\newif\ifsingle
\ifsingle
\documentclass[12pt,draftclsnofoot, onecolumn]{IEEEtran}		
\else		
\documentclass[10pt,final, twocolumn]{IEEEtran}
\fi

\usepackage{mathtools}
\usepackage{framed}
\usepackage{algpseudocode}
\usepackage{cite}
\usepackage{graphicx}
\usepackage{amsmath}
\usepackage{array}
\usepackage{blindtext}
\usepackage{epstopdf}
\usepackage{bm}
\usepackage{enumerate}
\usepackage{amsbsy}
\usepackage{amssymb}
\usepackage{amsthm}
\usepackage{amscd}
\usepackage{subfigure}
\usepackage{color}
\usepackage{booktabs}
\usepackage{multirow}
\usepackage{hhline}
\usepackage{makecell}
\usepackage{bbold}
\usepackage{url}
\usepackage{tabularx}
\usepackage{cellspace}
\usepackage{boldline}
\usepackage{balance}

\usepackage[ruled,linesnumbered]{algorithm2e}  %,algochapter
\SetKwInput{KwData}{\textbf{Input}} 
\usepackage{algpseudocode} 

\usepackage{dsfont}
\usepackage{xcolor}
\usepackage[bookmarks,colorlinks]{hyperref}

\def \bb           {\boldsymbol{b}}

\def \bh           {\boldsymbol{h}}

\def \bn           {\boldsymbol{n}}

\def \br           {\boldsymbol{r}}

\def \bx           {\boldsymbol{x}}
\def \by           {\boldsymbol{y}}
\def \bz           {\boldsymbol{z}}

\def \bA           {\boldsymbol{A}}
\def \bB           {\boldsymbol{B}}
\def \bC           {\boldsymbol{C}}

\def \bG           {\boldsymbol{G}}
\def \bH           {\boldsymbol{H}}
\def \bI           {\boldsymbol{I}}

\def \bP           {\boldsymbol{P}}
\def \bQ           {\boldsymbol{Q}}
\def \bR           {\boldsymbol{R}}
\def \bS           {\boldsymbol{S}}

\def \bW           {\boldsymbol{W}}

\def \bTheta       {\boldsymbol{\Theta}}

\def \bphi         {\boldsymbol{\phi}}

\def \calF         {\mathcal{F}}

\ifsingle
\newcommand{\figwidth}{0.65\columnwidth}

\else
\newcommand{\figwidth}{0.9\columnwidth}

\fi % ------------------------------------------

\begin{document}

%\title{Model-Based Deep Learning for Low-Resolution Signal Processing}
\title{LoRD-Net: Unfolded Deep Detection Network with Low-Resolution Receivers}

\author{
	\IEEEauthorblockN{Shahin Khobahi, Nir Shlezinger, 	Mojtaba Soltanalian, and Yonina C. Eldar\\
	} 
	\thanks{
	Parts of this work was presented in the IEEE International Conference on Acoustics, Speech, and Signal Processing (ICASSP), Brighton, United Kingdom, 2019 \cite{8683876}.
    This paper has received supports from the Benoziyo Endowment Fund for the Advancement of Science, the	Estate of Olga Klein -- Astrachan, the European Union’s Horizon 2020 research and innovation program under grant No. 646804-ERC-COG-BNYQ, from the Israel Science Foundation under grant No. 0100101, from the U.S. National Science Foundation under grant No. CCF-1704401, and from an Illinois Discovery Partners Institute (DPI) Seed Award.
		%	}			
		%	\thanks{
		S. Khobahi  and M. Soltanalian are with the ECE Dept., University of Illinois at Chicago, Chicago, IL  (e-mail: \{skhoba2, msol\}@uic.edu). 
		%	}
		%	\thanks{
        N. Shlezinger is with the School of ECE, Ben-Gurion University of the Negev, Be'er-Sheva, Israel (e-mail: nirshl@bgu.ac.il). 
         Y. C. Eldar is with the Faculty of Math and CS, Weizmann Institute of Science, Rehovot, Israel (e-mail: yonina.eldar@weizmann.ac.il). }

	\vspace{-1.0cm}
	
}

\maketitle
\pagestyle{plain}
\thispagestyle{plain}

\begin{abstract}
The need to recover high-dimensional signals from their noisy low-resolution quantized measurements is widely encountered in communications and sensing. In this paper,
we focus on the extreme case of one-bit quantizers, and propose a deep detector entitled  LoRD-Net for recovering information symbols from one-bit measurements. Our method is a model-aware data-driven architecture based on deep unfolding of first-order optimization iterations. LoRD-Net has a task-based architecture dedicated to recovering the underlying signal of interest from the one-bit noisy measurements without requiring prior knowledge of the channel matrix through which the one-bit measurements are obtained. The proposed deep detector has much fewer parameters compared to black-box deep networks due to the incorporation of domain-knowledge in the design of its architecture, allowing it to operate in a data-driven fashion while benefiting from the flexibility, versatility, and reliability of model-based optimization methods. LoRD-Net operates in a blind fashion, which requires addressing both the non-linear nature of the data-acquisition system as well as identifying a proper optimization objective for signal recovery. Accordingly, we propose a two-stage training method for LoRD-Net, in which the first stage is dedicated to identifying the proper form of the optimization process to unfold, while the latter trains the resulting model in an end-to-end manner. %The unfolding of a first-order optimization method for deriving the architecture of the UGD-Net allows for achieving a unified signal processing framework that paves the way for carrying out the signal recovery problem without an explicit knowledge of the channel matrix. 
We numerically evaluate the proposed receiver architecture for one-bit signal recovery in wireless communications and demonstrate that the proposed hybrid methodology outperforms both data-driven and model-based state-of-the-art methods, while utilizing small datasets, on the order of merely $\sim 500$ samples, for training. 
\end{abstract}

%\begin{IEEEkeywords}
%Low-resolution signal processing, MIMO systems,  deep unfolding, deep learning.
%\end{IEEEkeywords}

\vspace{-0.4cm}
\section{Introduction}
\label{sec:introduction}
\vspace{-0.1cm}
 
Analog-to-digital conversion plays an important role in digital signal processing systems. While physical signals take values in continuous-time over continuous sets, they must be represented using a finite number of bits in order to be processed in digital hardware \cite{eldar2015sampling}. This operation is carried out using analog-to-digital converters (ADCs), which typically perform uniform sampling followed by a uniform quantization of the discrete-time samples. 
When using high-resolution ADCs, this conversion induces a minimal distortion, allowing to effectively process the signal using methods derived assuming access to the continuous-amplitude samples.  
However, the cost, power consumption and memory requirements of ADCs grow with the sampling rate and the number of bits assigned to each sample \cite{walden1999analog}. Consequently, recent years have witnessed an increasing interest in digital signal processing systems operating with low-resolution ADCs. Particularly, in multiple-input multiple-output (MIMO) communication receivers, which are required to simultaneously capture multiple analog signals with high bandwidth,  there is a growing need to operate reliably with low-resolution ADCs \cite{andrews2014will}. The most coarse form of quantization is reduction of the signal to a single bit per sample, which may be accomplished via comparing the sample to some reference level, and recording whether the signal is above or below the reference. One-bit acquisition allows using high sampling rates
at a low cost and  low energy consumption. Due to such favorable properties of one-bit  ADCs, they have been employed in a wide array of applications, including in wireless communications \cite{jeon2018one, rao2020massive, 8683876}, radar signal processing \cite{ameri2019one, jin2020one,xi2020bilimo}, and sparse signal recovery \cite{xiao2019one, khobahi2019model}. 

The non-linear nature of low-resolution quantization makes symbol detection a challenging task. This situation is significantly exacerbated in practical one-bit communication and sensing where the channel is to be estimated in conjunction with symbol detection. A \emph{coherent} symbol detection task is concerned with recovering the underlying signal of interest from the one-bit measurements assuming the channel state information (CSI) is known at the receiver. On the other hand, the more difficult task of \emph{blind} symbol detection, which is the focus here, carries out recovery of the underlying transmitted symbols when CSI is not available.

Two main strategies have been proposed in the literature to facilitate operation with low-resolution ADCs: The first designs the overall acquisition system in light of the task for which the signals are acquired. For instance, MIMO communication receivers acquire their channel output in order to extract some underlying information, e.g., symbol detection. As the analog signals are not required to be recovered from their digital representation, one can design the  acquisition system to reliably infer the desired information while operating with low resolution ADCs \cite{shlezinger2018hardware, salamatian2019task, shlezinger2019deep, shlezinger2020learning,neuhaus2020task}. Such task-based quantization systems rely on pre-quantization processing, which requires dedicated hardware in the form of hybrid receiver architectures \cite{gong2019rf,ioushua2019family} or unique antenna structures \cite{wang2019dynamic,shlezinger2020dynamic}, which are configured along with the quantization rule. % in light of the underlying statistical model. 

An alternative approach to task-based quantization, which does not require additional configurable analog hardware and is the focus of the current work, is to recover the desired information from the distorted coarsely discretized representation of the signal in the digital domain. The main benefit of schemes carried out only in the digital domain is their simplicity of implementation, as they do not require to introduce modifications to the quantization system and circumvent the need for adding pre-quantization analog processing hardware. %Nonetheless, in order to deal with the statistical intricacies arising from low resolution quantization, model-based low-resolution signal processing schemes commonly assume a statistical model for the quantization error. A usual pick is that of an additive Gaussian noise, independent of the input, which stems from the Bussgang decomposition \cite{li2017channel,jacobsson2017throughput} and the additive quantization noise model of \cite{orhan2015low}.  Such strategies induce additional errors since the assumed model may not accurately reflect the distortion induced by quantization \cite{gray1998quantization}. 
In the context of MIMO systems, various methods have been proposed in the literature for channel estimation and signal decoding from quantized outputs, including model-based signal processing methods as surveyed in \cite{liu2019low}, as well as model-agnostic systems based on machine learning and data-driven techniques \cite{zhang2020deep, klautau2018detection,balevi2019one,balevi2019two,balevi2020autoencoder,kim2019machine,nguyen2020svm, nguyen2020linear}.

Most existing model-based detection algorithms require coherent operation, i.e., they rely on prior knowledge of the CSI and other system parameters. Among these works are the near-Maximum Likelihood (nML) detector proposed for one-bit MIMO receivers in \cite{choi2016near}, the linear receivers studied in \cite{risi2014massive, jacobsson2015one}, and the message passing based detectors considered in \cite{ivrlac2007mimo, mo2014channel}. The fact that such approaches require accurate CSI led to several works specifically dedicated to CSI estimation in the presence of low-resolution ADCs. These include  \cite{choi2016near, mezghani2018blind}, which studied maximum-likelihood estimation for recovering the CSI in the presence of one-bit data,  the works in \cite{li2017channel, jacobsson2017throughput}, which developed linear estimators for CSI estimation purposes in one-bit MIMO systems, and \cite{mo2017channel} which focuses on sparse channels and utilizes one-bit sparse recovery methods for CSI estimation. However, all these strategies inevitably induce non-negligible CSI estimation error, which may notably degrade the accuracy in signal detection based on the estimated CSI. %Furthermore, these aforementioned works deal with the channel estimation and the symbol detection as two disjoint tasks. This will be shown through the rest of this paper to be sub-optimal.

Over the past several years, data-driven methods, and specifically deep neural networks (DNNs), have attracted unprecedented attention from research communities across the board. The advent of low-cost specialized powerful computing resources and the continually increasing amount of massive data generated by the human population and machines, along with new optimization and learning methods, have paved the way for DNNs and machine learning-based models to prove their effectiveness in many engineering areas, such as computer vision and natural language processing \cite{lecun2015deep}. DNNs learn their mapping from data in a model-agnostic manner, and can thus facilitate non-coherent (blind) detection.

Previously proposed DNN-aided symbol detection techniques for communication receivers can be divided based on their receiver architectures; namely, those that utilize conventional machine learning architectures for detection, including \cite{farsad2017detection,corlay2018multilevel, liao2019deep}, and schemes combining DNNs with model-based detection methods, such as the blind DNN-aided receivers proposed in  \cite{shlezinger2019viterbinet,shlezinger2020deepsic,shlezinger2020data,he2019model} and the coherent detectors of \cite{samuel2019learning,takabe2019trainable}, see also surveys in \cite{balatsoukas2019deep,farsad2020data}. In the context of one-bit DNN-aided receivers, previous works to date focus mainly on the first approach, i.e., applying conventional DNNs for the overall detection task. 
Among these works are \cite{zhang2020deep, balevi2019two} and \cite{klautau2018detection}, which applied generic DNNs for channel estimation in one-bit MIMO receivers. The application of conventional architectures for symbol detection was studied in   \cite{balevi2019one,kim2019machine} and \cite{nguyen2020svm}, while  \cite{balevi2020autoencoder} showed that autoencoders can facilitate the  design of error correction codes for communications with one-bit receivers. Recently, the authors in \cite{nguyen2020linear} considered the problem of symbol detection for a one-bit massive MIMO system and proposed a linear estimator module based on the Bussgang decomposition technique combined with a model-driven neural network.

The vast majority of the aforementioned works on learning-aided one-bit receivers rely on conventional DNN architectures. Such DNNs require a massive amount of training samples and must be trained on data from the same (or a similar) statistical model as the one under which they are required to operate, imposing a major challenge in dynamic wireless communications. In fact, the use of generic black-box DNNs is mostly justified in applications where a satisfactory description of the underlying governing dynamics of the system is not achievable, as is the case in computer vision and natural language processing fields. As surveyed above, this is not the case in the field of one-bit MIMO systems. %There exists a large body of research around developing model-based signal processing techniques for symbol detection and channel estimation in such scenarios. 
This gives rise to the need that is bridging the gap between data-driven and model-based approaches in this context, and moving towards specialized deep learning models for signal processing techniques in one-bit MIMO systems---which is the aim of this work.

In this paper, we develop a hybrid model-based and data-driven system which learns to carry out blind symbol detection from one-bit measurements. %We initiate the work by assuming that the receiver has knowledge of the underlying model and its parameters, i.e., full CSI, and propose a deep network architecture to tackle the signal recovery problem and achieve high accuracy at affordable complexity. We  later show that the resulting system is in fact invariant of the channel parameterization, and thus can operate in a blind manner.
The proposed architecture, referred to as LoRD-Net (Low Resolution Detection Network), combines the well-established model-based maximum-likelihood estimator (MLE) with machine learning tools through the deep unfolding method \cite{hershey2014deep,monga2019algorithm, khobahi2020unfolded, agarwal2020deep, khobahi2020deep, naimipour2020upr} for designing DNNs based on model-based optimization algorithms. 
To derive LoRD-Net, we first formulate the MLE for the task of symbol detection from one-bit samples. Next, we resort to  first-order gradient-based methods for the MLE computation, and unfold the iterations onto layers of a DNN. The resulting LoRD-Net learns to carry out MLE-approaching symbol detection without requiring  CSI. 

Applying conventional gradient-based optimization methods requires  knowledge of the underlying system parameters, i.e., full CSI. % $\Theta=\{\bH, \bC\}$.
Hence, a typical approach to unfold such a symbol detection algorithm would be to estimate the unknown parameters from training, and substitute it into the unfolded network \cite{he2019model}. We show that instead of estimating the unknown system parameters, it is preferable to learn an \emph{alternative channel} which allows the receiver to detect the symbols reliably. Surprisingly, we demonstrate that the alternative channel learned by LoRD-Net is in general not the true channel. Based on this observation, we propose a two-stage training procedure, comprised of learning the proper optimization process to unfold, followed by an end-to-end training of the unfolded DNN. 

The proposed LoRD-Net has thus the following properties:
\begin{enumerate}
    \item[\emph{i)}] Compared to the vanilla MLE symbol detector, our model does not need to estimate the channel separately.
    \item[\emph{ii)}] Owing to its hybrid nature, it has low computational cost in operation  and is highly scalable, facilitating much faster inference as compared to its black-box data-driven and model-based counterparts.
    \item[\emph{iii)}] The proposed deep architecture is interpretable and has far fewer parameters compared to existing black-box deep learning solutions. This follows from the incorporation of domain-knowledge in the design of the network architecture (i.e., being model-based), allowing LoRD-Net to train with  much fewer labeled samples as compared to existing data-driven one-bit receivers.
\end{enumerate}%\emph{i)} compared to the vanilla MLE for symbol detection tasks, our model does not need to estimate the channel separately; \emph{ii)} due to its hybrid nature, it has a low computational cost in operation  and is highly scalable, allowing for a much faster inference as compared to its black-box data-driven and model-based counterparts; and \emph{iii)} the proposed deep architecture is interpretable and has much fewer parameters compared to the existing black-box deep learning solutions. This follows from the incorporation of domain-knowledge in the design of the architecture of the underlying network (i.e., being model-based), allowing UGD-Net to train with  much fewer labeled samples as compared to the existing data-driven one-bit receivers. %In addition, the proposed methodology allows for an incorporation of non-zero quantization thresholds at time of the quantization that further enable us to recover the amplitude information as well. 
We verify the above characteristics of LoRD-Net in an experimental study, where we show that training of the proposed LoRD-Net architecture can be performed with far fewer samples as compared to its data-driven counterparts, and demonstrate substantially superior performance compared to existing model-based and data-driven algorithms for symbol detection in massive MIMO channels with one-bit ADCs.

%To the best of our knowledge, there is no existing work that considers the deep unfolded networks in a one-bit wireless communication system. Therefore, the motivation behind our work is to bridge the gap between well-established signal processing techniques and modern deep learning models in the context of data detection and channel estimation in a one-bit MIMO system. In particular, we take the well-established iterative approaches, specifically developed for low-resolution signal processing and inference in one-bit MIMO systems, and use the deep unfolding technique to derive a model-based deep architecture specifically tailored for the task of symbol detection and channel estimation. Next, we make use of deep learning tools to boost the performance of the underlying inference optimization algorithm in terms of speed of convergence and effectiveness. The resulting hybrid model differ from the existing deep learning based methods in that it leads to novel model-based \emph{deep architectures} with novel activation functions specifically designed to resemble the iterations of a well-established optimization algorithm, and that it incorporates both parameterized and non-parameterized mathematical models for such a complex systems. Note that \emph{since the emerging deep architecture is sparser due to incorporation of problem-level reasoning, the training of the proposed architecture will be less data hungry and much faster than general black-box generic DNNs---thus, paving the way for real-time inference}.

% \vspace{3pt}

The rest of the paper is organized as follows. In Section \ref{sec:SystemModel}, we present the considered system model and the corresponding MLE formulation. In Section \ref{sec:Proposed}, we derive LoRD-Net by unfolding the first-order gradient iterations associated with the MLE computation, and present its two-stage training procedure. %We conclude Sec. \ref{sec:Proposed} by providing the training and inference procedure of the proposed UGD-Net. 
Section \ref{sec:NumericalStudy} provides a detailed numerical analysis of LoRD-Net applied to MIMO communications. %, and compare its performance with the existing model-based and data-driven methodologies. 
Finally, Section~\ref{sec:Conclusion} concludes the paper. 
 
%  \vspace{3pt}
 
Throughout the paper, we use the following notation. Bold lowercase and bold uppercase letters denote vectors and matrices, respectively. We use $(\cdot)^T$, $\mathrm{Diag}(\cdot)$, and $\mathrm{sign}(\cdot)$, and $\mathrm{log}\{\cdot\}$  to  denote the transpose operator, the diagonal matrix formed by the entries of the vector argument,  the sign operator, and the natural logarithm, respectively. The symbol $\odot$ represents the Hadamard product, while $\mathbf{1}$ and $\mathbf{0}$ are the all-one and all-zero vectors/matrices. %and $\mathrm{Diag}(\cdot)$ denotes the diagonal matrix formed by the entries of the vector argument. The function $\mathrm{sign}(\cdot)$ denotes the one-bit sampler defined as $\mathrm{sign}(x)=+1$ if $x\geq 0$ and $\mathrm{sign}(x)=-1$ if $x< 0$. 
 The $i$-th entry of the vector $\bx$ is $x_i$, and $\|\bx\|_p$ is the $\ell_p$-norm of $\bx$; % the vector argument and is defined as $(\sum_i|x_i|^p)^{\frac{1}{p}}$. 
%Moreover, $\mathrm{log}\{\cdot\}$ denotes the natural logarithm, and $[\cdot]_{-}$ is an operator retaining the negative elements of the vector argument. 
 $\mathcal{M}^n$ is the $n$-ary Cartesian product of a set $\mathcal{M}$, and $\mathds{S}_{+}$ denotes the cone of symmetric positive definite matrices.

%\nocite{khobahi2018signal, shlezinger2018hardware, salamatian2019task, shlezinger2019deep, shlezinger2020learning} \nocite{gong2019rf, mendez2016hybrid}  \nocite{wang2019dynamic}
 
 \vspace{-0.2cm}
\section{System Model and Preliminaries}
\label{sec:SystemModel}
\vspace{-0.1cm}
In this section, we discuss the considered system model. We focus on one-bit data acquisition and blind signal recovery. We then formulate the MLE for this problem, which is used in designing the LoRD-Net architecture in Section~\ref{sec:Proposed}.

\vspace{-0.2cm}
\subsection{Problem Formulation}
\label{subsec:Model}
\begin{figure}
    \centering
    \includegraphics[width=\columnwidth]{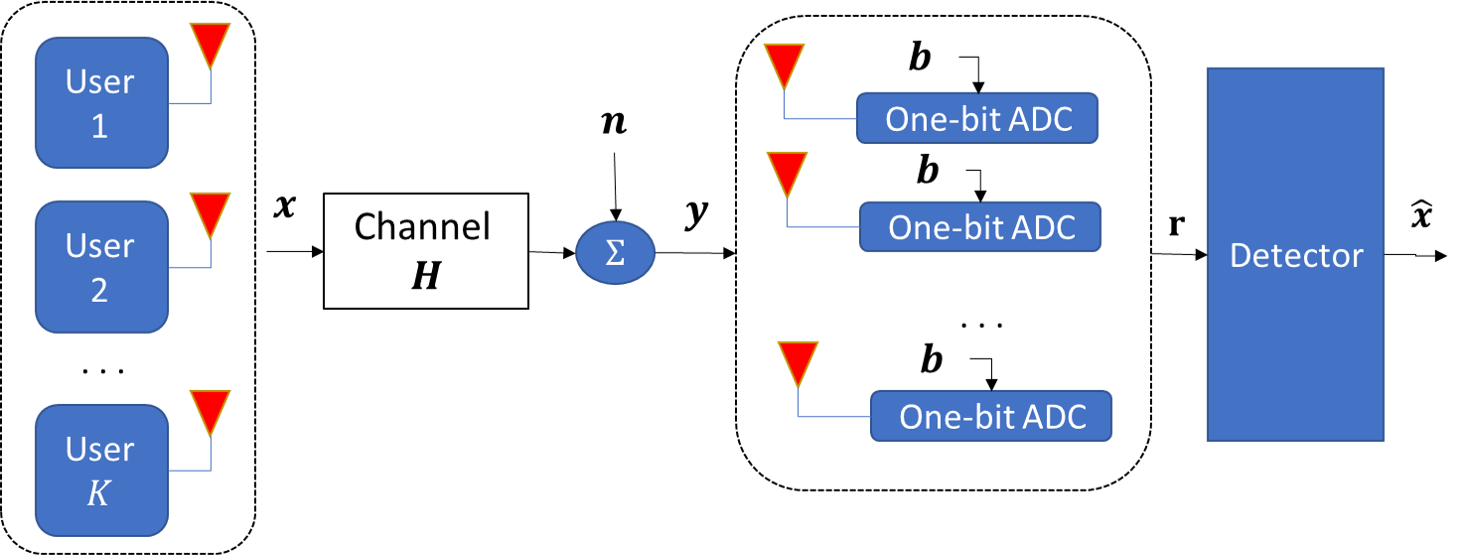}
    \caption{System model illustration.}
    \label{fig:SystemModel}
\end{figure}

\vspace{-0.1cm}
 We consider a low-resolution data-acquisition system which utilizes $m$ one-bit ADCs. By letting  $\by\in\mathds{R}^m$ denote the received signal, the discrete output of the ADCs can be written as $\br = \mathrm{sign}\left(\by - \bb\right)$,  
where $\bb\in\mathds{R}^m$ denotes the vector of quantization thresholds, and $\mathrm{sign}(\cdot)$ is the sign function, i.e., $\mathrm{sign}(x) = +1$ if $x\geq 0$ and $\mathrm{sign}(x)=-1$ otherwise. The received vector $\by$ is statistically related to  the unknown vector of interest $\bx\in\mathcal{M}^n\subseteq\mathds{R}^n$ according to the following linear relationship:
\begin{align}\label{eq:3}
    \by = \bH\bx + \bn,
\end{align} 
where $\bn\sim\mathcal{N}(\mathbf{0}, \bC)$ denotes additive Gaussian noise with a covariance matrix of the form $\bC = {\bf{\rm Diag}}(\sigma_0^2, \sigma_1^2, \dots, \sigma_{m-1}^2)$ with diagonal entries $\{\sigma_i^2\}_{i=0}^{m-1}$ representing the noise variance at each respective dimension, and $\bH\in\mathds{R}^{m\times n}$ is the channel matrix.  We assume that the elements of the unknown vector $\bx$ are chosen independently from a finite alphabet $\mathcal{M}=\{s_1, s_2, \cdots, s_{|\mathcal{M}|}\}$. This setup represents low-resolution receivers in uplink multi-user MIMO systems, where  $\bx$ is the symbols transmitted by the users, and $\by$ is the corresponding channel output, as illustrated in Fig.~\ref{fig:SystemModel}.

The overall dynamics of the system are thus  compactly expressed as:
\begin{align}\label{eq:sysmodel}
{\br} = \mathrm{sign}(\bH\bx + \bn - \bb).
\end{align}
%Throughout this work, we assume the quantization thresholds vector $\bb$ is known.
In the sequel, we refer to $\Theta=\{\bH, \bC\}$ as the system parameters. Note that the above system model can be modified using conventional transformations to accommodate a complex-valued system model.

Our main goal is to perform the task of symbol detection, i.e., recover $\bx$, from the collected one-bit measurements $\br$. We focus on blind (non-coherent) recovery, namely, the system parameters $\Theta=\{\bH, \bC\}$, i.e., the channel matrix and the covariance of the noise, are not available to the receiver. Nonetheless, the receiver has access to a limited set of $B$  labeled samples $\{\bx_p^b, \br_{p}^b\}_{b=0}^{B-1}$, representing, e.g., pilot transmissions. The  quantization thresholds of the ADCs, i.e., the vector $\bb$, are assumed to be fixed and known. While we do not consider the selection of $\bb$ in the following, we discuss in the sequel how its optimization can be incorporated into the detection method.

\vspace{-0.2cm}
\subsection{Maximum Likelihood Recovery}
\label{subsec:MLE}
\vspace{-0.1cm}
To understand the challenges associated with blind low-resolution detection, we next discuss the MLE for recovering $\bx$ from $\br$. In particular, the intuitive model-based approach is to utilize the labeled data to estimate the system parameters $\Theta$, and then to use this estimation to compute the coherent (non-blind) MLE. Therefore, to highlight the limitations of this strategy,  we assume here that the system parameters $\Theta = \{\bH, \bC\}$ are fully known at the receiver. Let
\begin{align}
    \mathcal{F}_{\Theta}(\bx; \br) & \triangleq \mathrm{log}\,\mathrm{Pr}(\br|\bx,\Theta) \notag \\
    &\stackrel{(a)} = -\sum_{i=0}^{m-1}\mathrm{log} \left\{ Q \left( \frac{r_i}{\sigma_i} \left( b_i - \bh_i^T\bx  \right) \right)\right\}, \label{eq:MLObj}
\end{align}
represent the log-likelihood objective for a given vector of one-bit observations $\br$, where $(a)$ is proven in \cite{8683876}. The coherent MLE is then given by
\begin{align}\label{eq:coherentMLE}
 \hat{\bx}_{\mathrm{ML}}(\br) = \underset{\bx\in\mathcal{M}^n}{\mathrm{argmax}}\;\;\mathcal{F}_{\Theta}(\bx;\br).
\end{align}

Although the MLE in \eqref{eq:coherentMLE} has full accurate knowledge of the parameters $\Theta$, its computation is still challenging. The main difficulty  emanates from solving the underlying optimization problem in the discrete domain, implying that the MLE requires an exhaustive search over the discrete domain $\mathcal{M}^n$, whose computational complexity grows exponentially with $n$. 
A common strategy to tackle the discrete optimization problem in \eqref{eq:coherentMLE} is to relax the search space to be continuous. This results in the following relaxed unconstrained MLE rule:
\begin{align}\label{eq:relaxed}
    \bar{\bx}_{\Theta}(\br) = \underset{\bx\in\mathds{R}^n}{\mathrm{argmax}}\;\;\mathcal{F}_{\Theta}(\bx;\br).
\end{align}
The optimization problem in \eqref{eq:relaxed} is convex due to the  log-concavity of $Q(\cdot)$, and thus can be solved using first-order gradient optimization. In particular, the gradient of the negative log-likelihood function with respect to the unknown vector $\bx$ can be compactly expressed as \cite{8683876}: 
\begin{eqnarray}
    \nabla_{\bx}\mathcal{F}_{\Theta}(\bx;\br) = \bH^T\tilde{\bR}\;\boldsymbol\eta \left(\tilde{\bR}\left(\bb - \bH\bx\right) \right),
    \label{eqn:gradient1}
\end{eqnarray}
where $\boldsymbol{\eta}$ is a non-linear function defined as $\boldsymbol\eta(\bx) \triangleq Q^{'}(\bx) \oslash  Q(\bx)$, in which the operator $\oslash$ denotes the element-wise division operation, $Q^{'}(x)$ is the derivative of $Q(x)$, that is given by the negative probability density function of a standard Normal distribution, and $\tilde\bR = \bR\bC^{-\frac{1}{2}}$ is the semi-whitened version of the \emph{one-bit matrix} $\bR={\rm Diag}\left(r_0,\dots, r_{m-1}\right)$.

As $ \bar{\bx}_{\Theta}(\br)$ obtained via \eqref{eq:relaxed} is not guaranteed to take values in $\mathcal{M}^n$, the final estimate of the symbols is obtained by applying a projection operator $\mathcal{P}_{\mathcal{M}^n}: \mathds{R}^n\mapsto\mathcal{M}^n$ to $\bar{\bx}(\br)$. This operator maps the continuous input vector onto its closest lattice point on the discrete set $\mathcal{M}^n$, i.e., 
\begin{align}\label{eq:projection}
    \mathcal{P}_{\mathcal{M}^n}(\bx) = \underset{\bz\in\mathcal{M}^n}{\mathrm{argmin}} \|\bz - \bx\|_2^2.
\end{align}

Tackling a discrete program via continuous relaxation, as done in \eqref{eq:relaxed}, is subject to an inherent drawback. As a case in point, one can only expect $\bar{\bx}_{\Theta}(\br)$ to provide an accurate approximation of the true MLE if the real-valued vector $\bar{\bx}_{\Theta}(\br)$ is very close to the discrete valued MLE $\hat{\bx}_{\rm{ML}}(\br)$. In such a case, the MLE  is obtained by projecting into the lattice points in $\mathcal{M}^n$. However, this is not the case in many scenarios, and specifically, when the noise variance in each respective dimension is high. In other words, it is not necessarily the case that the minimizer of the objective function on the continuous domain \eqref{eq:relaxed} is close to the MLE, which takes values in the discrete set $\mathcal{M}^n$. 
%Specifically, if one is to tackle the discrete optimization problem in \eqref{eq:blind0} or \eqref{eq:coherentMLE} via continuous relaxation, the fixed likelihood function obtained via an estimation of the unknown system parameters or the true ones are usually the optimal choice for such a relaxation. Specifically, consider the simpler case of coherent detection.
Note that utilizing the true system parameters will only lead to optimal estimates when considering the original discrete problem \eqref{eq:coherentMLE}. In fact, one can no longer necessarily argue that the true system parameters are optimal choices for $\Theta$ in the  relaxed MLE. This insight, which is obtained from the computation of the coherent MLE, is used in our derivation of the blind unfolded detector in the following section.

\vspace{-0.2cm}
\section{Proposed Methodology}
\label{sec:Proposed}
\vspace{-0.1cm}
In this section, we present the proposed \textbf{Lo}w \textbf{R}esolution \textbf{D}etection \textbf{Net}work, abbreviated as \textbf{LoRD-Net}. We begin with a high-level description of LoRD-Net in Subsection~\ref{subsec:Description}. Then, we present the unfolded architecture in Subsection~\ref{subsec:Architecture} and discuss the training procedure in Subsection~\ref{subsec:train}. Finally, we provide a discussion in Subsection~\ref{subsec:discussion}.

\vspace{-0.2cm}
\subsection{High-Level Description}
\label{subsec:Description}
\vspace{-0.1cm}
As noted in the previous section, the intuitive approach to blind symbol detection is to utilize the labeled data  $\{\bx_p^b, \br_{p}^b\}_{b=0}^{B-1}$ to estimate the true system model $\Theta$, and then to recover the symbol vector $\bx$ from $\br$ using the MLE. Nonetheless, the coherent MLE \eqref{eq:coherentMLE} is computationally prohibitive, while its relaxed version in \eqref{eq:relaxed} may be inaccurate. Alternatively, one can seek a purely data-driven strategy, using the data to train a black-box highly-parameterized DNN for detection, requiring a massive amount of labeled samples. Consequently, to facilitate accurate detection at affordable complexity and with limited data, we design LoRD-Net via model-based deep learning \cite{shlezinger2020model}, by combining the learning of a {\em competitive objective}, combined with {\em deep unfolding} of the relaxed MLE. 

\begin{figure}
	\centering
	\includegraphics[width =\linewidth]{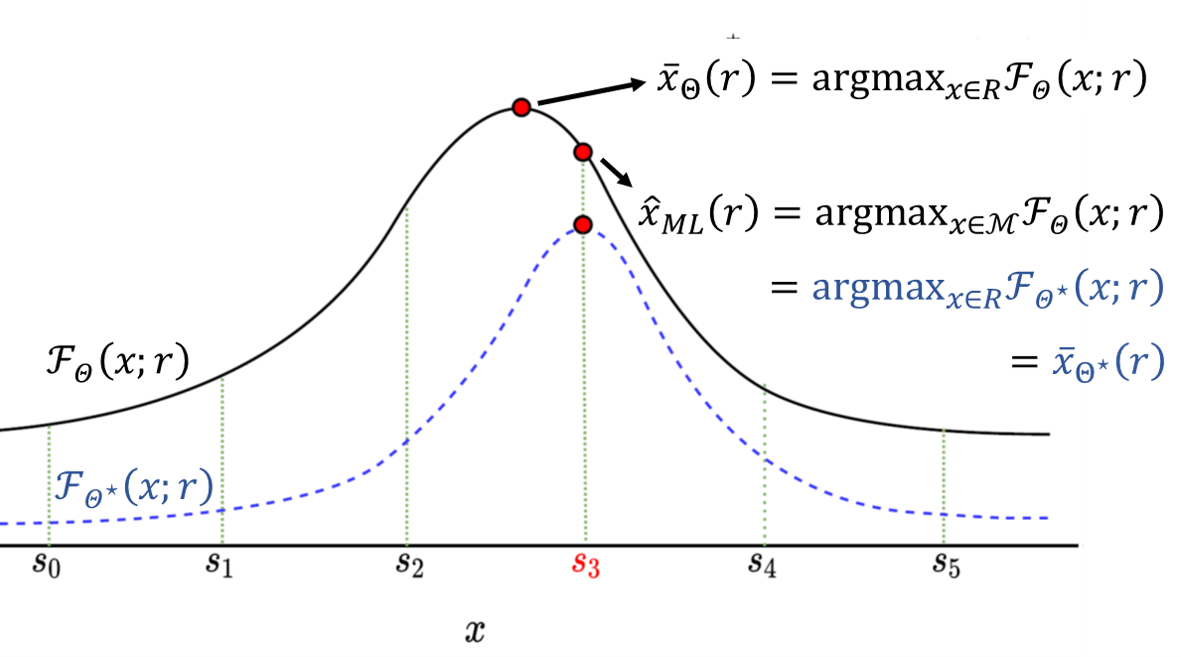}
	\vspace{-.8cm}\caption{An illustration of the relation between the optimal point of a competitive objective function (dashed blue line) and the true MLE $\hat{\bx}_{\rm ML}$ obtained by an exact maximization of the log-likelihood objective function (solid black line) over the discrete set $\mathcal{M}$ as well as an approximation of the MLE $\bar{\bx}_{\Theta}$ obtained by a maximization of the log-likelihood objective function over the continuous space $\mathds{R}$, when the true transmitted symbol is $s_3\in\mathcal{M}$.}
	\label{fig:surrogate}
	\vspace{4pt}
\end{figure}

Learning a competitive objective refers to the setting of the unknown system parameters $\Theta$. However, the goal here is not to estimate the {\em true system parameters}, but rather the ones for which the solution to the relaxed MLE coincides with the true value of $\bx$. This system identification problem can be written as
\begin{align}\label{eq:SIP}
  \!  \mathcal{F}_{\Theta^{\star}}\!(\br;\bx)\! =\! \underset{\Theta}{\mathrm{min}}~
    \frac{1}{B}\sum_{b=0}^{B-1}\left\| \bar{\bx}_{\Theta}(\br_p^b)  \! -\! \bx_p^b
    \right\|_2^2,
\end{align}
where $\bar{\bx}_{\Theta}$ is the relaxed MLE \eqref{eq:relaxed}. 
The optimization problem \eqref{eq:SIP} yields a surrogate  objective function $\calF_{\Theta^{\star}}$, or equivalently, a set of system parameters $\Theta^{\star}$, referred to as a {\em competitive objective} to the true $\calF_{\Theta}$. An illustration of such a competitive objective obtained for the  case of $n=1$ is depicted in Fig.~\ref{fig:surrogate}.

The main difficulty in solving \eqref{eq:SIP} stems from the fact that 
 $\bar{\bx}_{\Theta}(\br) = \underset{\bx\in\mathds{R}^n}{\mathrm{argmax}}\;\;\mathcal{F}_{\Theta}(\bx;\br)$ is not differentiable with respect to the system parameters $\Theta$. We overcome this obstacle by applying a differentiable approximation of $\bar{\bx}(\br)$, or equivalently, an algorithm that approximates the $\mathrm{argmax}$ operator specific to our problem. Since   $\bar{\bx}_{\Theta}(\br)$ can be computed by first-order gradient methods, we design a deep unfolded network \cite{monga2019algorithm} to compute the relaxed MLE in manner which is differentiable with respect to $\Theta$. The usage of deep unfolding allows not only to learn a competitive objective via \eqref{eq:SIP}, but also results in accurate inference with a reduced number of iterations compared to model-based first-order gradient optimization. Furthermore, the unfolded network utilizes a relatively small amount of trainable parameters, thus enabling learning from small amounts of labeled samples.

 \vspace{-0.2cm}
\subsection{LoRD-Net Architecture}
\label{subsec:Architecture}
\vspace{-0.1cm}
We now present the architecture of LoRD-Net, which maps the low resolution $\br$ into an estimated $\hat{\bx}$. For given system parameters $\Theta$ whose learning is detailed in Subsection~\ref{subsec:train} based on the competitive objective rationale described above, LoRD-Net is obtained by unfolding the iterations of a first-order optimization of the relaxed MLE \eqref{eq:relaxed}.  
 Our derivation thus begins by formulating the first-order methods to iteratively solve \eqref{eq:relaxed} for a given $\Theta$. 
 
 Let $g_{\phi_i}:\mathds{R}^n \mapsto \mathds{R}$ be a parametrized operator defined as $g_{\phi_i}(\bx;\Theta,\br) = \bx - \bG_i\nabla_{\bx} \mathcal{F}_{\Theta}(\bx; \br)$, where $\bG_i\in\mathds{R}^{n\times n}$ is a positive-definite weight matrix and $\phi_i = \{\bG_i\}$ denotes the set of parameters of the operator~$g_{\phi_i}$. Such a linear operator can be used to model a first-order optimization solver by considering a composition of $t$ mappings of the form:
\begin{align}
    \label{eq:mapping}
    \bx_{t+1} = \mathcal{G}^t_{\bphi}(\bx_0; \Theta, \br) 
    &= \bx_t - \bG_t\nabla_{\bx} \mathcal{F}_{\Theta}(\bx_t; \Theta,\br), \\ &\triangleq g_{\phi_t} \circ \cdots \circ g_{\phi_{1}} \circ g_{\phi_0}(\bx_0; \Theta,\br)\notag 
\end{align}
where $\bx_0$ is an initial point, $\bphi = \{\phi_0, \cdots, \phi_{t-1}\}$ is the set of parameters of the overall mapping $\mathcal{G}_{\bphi}^t$.   The mapping \eqref{eq:mapping} is differentiable with respect to the system parameters $\Theta$, and its local weights $\bphi$. For a fixed number of iterations $L$, the resulting function $\mathcal{G}_{\bphi}^L(\bx_0; \Theta, \br)$ is thus differentiable with respect to the set of parameters $\{\bphi, \Theta\}$ and its input (unlike the original $\rm{argmax}$ operator). Therefore, it can now be used as a differentiable approximation of $\bar{\bx}_{\Theta}(\br)$, which allows for a training (optimization) over the set of its parameters based on the gradient-based training algorithms and the back-propagation technique.

%Compared to model-based optimization methods from which UGD-Net will emerge, we will see that UGD-Net tends to require less iterations, i.e., layers, and typically carries out the inference task quicker, once trained.  Hence, it can be interpreted as a model-based DNN which is faithful to the original system model, giving rise to an \emph{interpretable} deep learning model benefiting from the best of both worlds of data-driven and model-based methodologies. In the  the following, we build the UGD-Net based upon the functions of the form \eqref{eq:mapping}. Next, we make use of the derived architecture to further elaborate on how $\Theta$ and $\bphi$ can be inferred and learned to maximize symbol detection accuracy, as part of the training procedure in Sub-Section \ref{subsec:train}.

\begin{figure*}
    \centering
    \includegraphics[width=\linewidth]{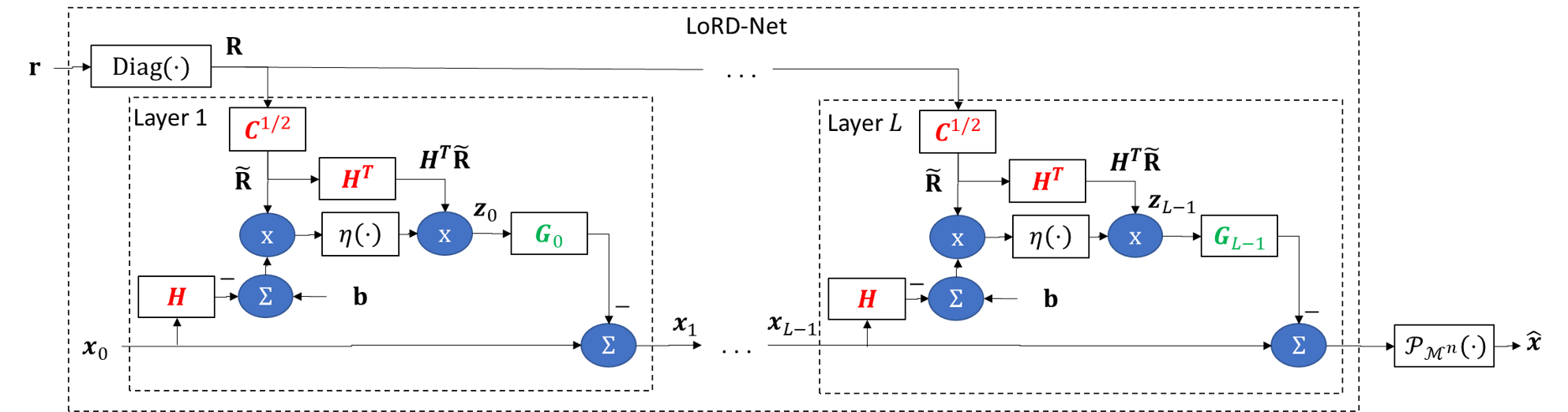}
    \caption{An illustration of LoRD-Net, where trainable system parameters and unfolded weights are highlighted in red and green colors, respectively.}
    \label{fig:UGD_Net1}
\end{figure*}

Following the deep unfolding framework \cite{monga2019algorithm}, the  function $\mathcal{G}
_{\bphi}^L(\bx_0; \Theta, \br)$ can be implemented as a $L$-layer feed-forward neural network, where the initial point $\bx_0$ and the one-bit samples $\br$ constitute the input to the network, and with trainable parameters that are given by $\{\Theta, \bphi\}$. 
By \eqref{eqn:gradient1}, the $i$-th layer computes:%. From an initial point $\bx_0$, every layer $i\in\{0,1,\cdots,L-1\}$ of the UGD-Net is tuned to compute:
\begin{align} %\label{eq:objective_grad}
    g_{\phi_i}(\bx_i;\Theta, \br)= \bx_i - \bG_i\bz_i\label{eq:proposed1}, \text{ with}\\
    \bz_i =  \bH^T\tilde{\bR}\;\boldsymbol\eta \left(\tilde{\bR}\left(\bb - \bH\bx_i\right) \right),
\end{align}
where the overall dynamics of the LoRD-Net is given by:
\begin{align}
 \mathcal{G}_{\bphi}^{L}(\bx_0; \Theta, \br) \!=\! g_{\phi_{L\!-\!1}} \circ g_{\phi_{L\!-\!2}} \circ \cdots \circ g_{\phi_{0}}(\bx_0; \Theta, \br).\label{eq:decodermodule}
\end{align}
% \begin{oframed}
% \vspace{-0.3cm}
% \label{UGD-Net}
% \noindent \\\textbf{UGD-Net Computation Dynamics:} \\
% Initialize \mbox{$\bx_0$}. \\ Every layer $i\in\{0,1,\ldots,L-1\}$ is tuned to compute: 
% \begin{equation}\label{eq:objective_grad}
%     \bz_i =  \bH^T\tilde{\bR}\;\boldsymbol\eta \left(\tilde{\bR}\left(\bb - \bH\bx_i\right) \right),
% \end{equation}
% and 
% \begin{equation} 
% g_{\phi_i}(\bx_i;\Theta, \br)= \bx_i - \bG_i\bz_i.\label{eq:proposed1}
% \end{equation}
% The overall dynamics of the UGD-Net is given by:
% \begin{align}
%  \mathcal{G}_{\bphi}^{L}(\bx_0; \Theta, \br) \!=\! g_{\phi_{L\!-\!1}} \circ g_{\phi_{L\!-\!2}} \circ \cdots \circ g_{\phi_{0}}(\bx_0; \Theta, \br).\label{eq:decodermodule}
% \end{align}
% \vspace{-0.2cm}
% \end{oframed}
Each vector $\bx_i$ in \eqref{eq:proposed1} represents the input to the $i$-th layer (or equivalently, the output of the previous iteration), with $\bx_0$ being the input of the entire network (which represents the initial point for the optimization task). Upon the arrival of any new one-bit measurement $\br$, the recovered symbols $\hat{\bx}$ are obtained by feed-forwarding $\br$ through the $L$ layers of LoRD-Net. In order to obtain discrete samples, the output of LoRD-Net is projected into the feasible discrete set $\mathcal{M}^n$, viz.
\begin{align}
    \hat{\bx} = \mathcal{P}_{\mathcal{M}^n}\left(\mathcal{G}^{L}_{\bphi}(\bx_0;\Theta,\br)\right).
\end{align}
%When the network is properly trained, UGD-Net is expected to carry out learned and accelerated first-order optimization, tuned to operate even in channel conditions for which such an approach does not yields the MLE for the true channel. 
An illustration of LoRD-Net is depicted in Fig.~\ref{fig:UGD_Net1}.

We note that one can also propose an alternative architecture, derived by applying the projection operator $\mathcal{P}_{\mathcal{M}^n}$ at the output of each layer, i.e., by defining $g_{\phi_i}(\bx_i;\theta,\br) = \mathcal{P}_{\mathcal{M}^n}(\bx_i - \bG_i\bz_i)$. Such a setting corresponds to the unfolding of a projected gradient descent method. However, our numerical investigations have consistently shown that such an architecture suffers from the vanishing gradient problem during training and a significant degradation in performance. As a result, we implement LoRD-Net while applying the projection operator once on the output of the network, and only during inference, as discussed above.

In principle, one can fix  $\bG_i = \delta \bI$ for some $\delta > 0$, for which \eqref{eq:decodermodule} represents $L$ steps of gradient descent with step size $\delta$. In the unfolded implementation, the weights $\{\bG_i\}$ are tuned from data, allowing to detect with less iterations, i.e., layers. As a result, once LoRD-Net is trained, i.e., its weight matrices $\bphi = \{\bG_i\}$ and the unknown system parameters $\Theta$ are learned from data, it is capable of carrying out fast inference, owing to its hybrid model-based/data-driven structure. Furthermore, the number of iterations $L$ is optimized to boost fast inference  in the training procedure, as detailed in the following.

\vspace{-0.2cm}
\subsection{Training Procedure}
\label{subsec:train}
\vspace{-0.1cm}
Herein, we present the training procedure for LoRD-Net. In particular, our main goal is to perform inference of the unknown system parameters $\Theta$ based on the rationale detailed in Subsection~\ref{subsec:Description}, i.e., to obtain a competitive objective. The learning competitive objective is used to tune the weights of the unfolded network $\bphi$. Accordingly, we present a two-stage training procedure for  LoRD-Net \eqref{eq:decodermodule}. Once the training of the LoRD-Net is completed, it carries out symbol detection from one-bit information without requiring the knowledge of system parameters $\Theta$. %RED
 \subsubsection{Training Stage 1 - Learning a Competitive Objective}  
% \vspace{2pt}
% $\bullet$ \emph{Training Stage 1 (RSIP)- Learning the System Parameters:}
%In this part, we present the first stage of the training process for the proposed UGD-Net. 
The first stage corresponds to learning the unknown system parameter $\Theta$. However, as formulated in \eqref{eq:SIP}, we do not seek to estimate the true values of the channel matrix $\bH$ and noise covariance $\bC$, but rather learn the surrogate values which will facilitate accurate detection using the relaxed MLE formulation. We do this by taking advantage of two propertities of LoRD-Net: The first is the differentiability of the unfolded architecture with respect to $\bTheta$, which facilitates gradient-based optimization optimization; The second is the fact that for  $\bG_i = \delta \bI$, LoRD-Net essentially implements $L$ steps of gradient descent with step size $\delta$ over the convex objective \eqref{eq:relaxed}, and is thus expected to reach its maxima.

Based on the above properties, we fix a relatively large number of layers/iterations $L$ for this training stage, and fix the weights $\bphi$ to $\bG_i  = \delta \bI$. Under this setting, the output of LoRD-Net $\mathcal{G}^L_{\bphi=\{\delta\bI\}}(\bx;\Theta,\br)$ represents an approximation of the relaxed MLE for a given parameter $\Theta$, denoted $\bar{\bx}_{\Theta}(\br)$, i.e., we have that 
\begin{align}\label{eq:UGDNetapprox}
   \bar{\bx}_{\Theta}(\br)\approx \mathcal{G}^L_{\bphi  =\{\delta\bI\}}(\bx_0;\Theta,\br).
\end{align}
We refer to the setting $\bphi=\{\delta\bI\}$ using in this stage as the \emph{basic} optimization policy. Note that as the number of layers grows large, the above approximation becomes more accurate. Hence, by substituting \eqref{eq:UGDNetapprox} into \eqref{eq:SIP} and replacing $\bar{\bx}_{\Theta}(\br_p^{i})$ with the corresponding outputs of LoRD-Net, we formulate the loss measure of the first training stage of LoRD-Net as:
\begin{align}\label{eq:training1}
    \Theta^\star = \underset{\Theta  }{\mathrm{argmin}}\;\;\frac{1}{B}\sum_{i=0}^{B-1}\left\|\mathcal{G}^{L}_{\bphi=\{\delta\bI\}}(\bx_0;\Theta,\br_p^i) - \bx_p^i\right\|_2^2. 
\end{align}
Owing to the differentiable nature of $ \mathcal{G}^L_{\bphi }(\bx_0;\Theta,\br)$ with respect to $\Theta$, we recover $ \Theta^\star$ based on \eqref{eq:training1} using conventional gradient-based training, e.g., stochastic gradient descent with backpropagation, as detailed in our numerical evaluations description in Section~\ref{sec:NumericalStudy}

\subsubsection{Training Stage 2 - Learning the Unfolded Weights}
Having learned the unknown system parameters $\Theta$ in Stage 1,  we turn to tuning the parameters of LoRD-Net, i.e., the set $\bphi = \{\bG_i\}$. We note that in Stage 1, the rationale was to use the basic optimization policy $\bphi=\{\bG_i = \delta \bI\}_{i=0}^{L-1}$ with a large number of layers $L$, exploiting the insight that under this setting, LoRD-Net effectively implements conventional gradient descent. However, once Stage~1 is concluded and  $\Theta^\star$ is learned, it is preferable to reduce the number of layers $L$ compared to that used in Stage~1, thus exploiting the ability of the unfolded network to carry out faster inference compared to their model-based iterative counterparts by learning the weights applied in each iteration \cite{gregor2010learning,monga2019algorithm}. Consequently, the first step in this stage is to set a number of layers to a value which can potentially be smaller than that used in the first training stage, and then optimize the weights according to the following criterion: 
\begin{align}\label{eq:training2emprical}
\!   \bphi^\star\!=\! \underset{\bphi}{\mathrm{argmin}}\;\frac{1}{B}\sum_{i=0}^{B-1}\left\|\mathcal{G}^{L}_{\bphi=\{\bG_l\}_{l=1}^L}\!(\bx_0;\Theta^\star,\br_p^i) \!-\! \bx_p^i\right\|_2^2.
\end{align}

Generally speaking, in order for a first-order optimizer (LoRD-Net in this case) to provide a descent direction at each iteration (layer), the pre-conditioning matrices must be positive-semidefinite so that each iteration does not reverse the gradient direction. To incorporate this requirement into LoRD-Net training, we  re-parameterize the pre-conditioning matrices by writing $\{\bG_i = \bW_i\bW_i^T\}$ and performing the traning over the matrices $\{\bW_i\}$. The resulting two-stage training algorithm is summarized as Algorithm~\ref{alg:Algo1}.
%In this work, we consider a diagonal structure on the pre-conditioning matrices, i.e., we set $\bW_i = \mathrm{Diag}(\delta_i^0, \delta_i^1, \cdots, \delta_i^{n-1})$ which accounts for a total of $nL$ trainable parameters.
%\vspace{4pt} 

\begin{algorithm}  
	\caption{Training LoRD-Net}
	\label{alg:Algo1}
	\KwData{Labeled data $\{\bx_p^b, \br_{p}^b\}_{b=0}^B$ }
	{\bf Stage 1 Init:} Fix (large) $L$, step-size $\delta\in(0,1)$, and weights $\bG_l=\delta\bI$. 
	Initialize \mbox{$\bx_0$} \;
	Optimize $\Theta^\star$ via \eqref{eq:training1} \tcp*{Stage 1}
	{\bf Stage 2 Init:} Fix (small) $L$.
	Initialize \mbox{$\bx_0$} \;
	Set the trainable parameters to $\{\bG_i=\bW_i\bW_i^T\}$ \;
    Optimize $\bphi^\star$  according to \eqref{eq:training2emprical} \tcp*{Stage 2}
	\KwOut{LoRD-Net parameters $\{\Theta^\star,\bphi^\star\}$ }
\end{algorithm}

When the network is properly trained, LoRD-Net is expected to carry out learned and accelerated first-order optimization, tuned to operate even in channel conditions for which such an approach does not yield the MLE for the true channel.

\vspace{-0.2cm}
\subsection{Discussion}
\label{subsec:discussion}
\vspace{-0.1cm}
LoRD-Net is a data-driven  acquisition system based on unfolding first-order gradient optimization methods, designed for low-resolution MIMO receivers operating without analog processing. 
Its model-awareness enables the receiver to learn to accurately infer from smaller training sets compared to conventional DNN architectures applied to such setups, as suggested, e.g., in \cite{balevi2019one}, giving rise to the possibility of tracking block-fading channel conditions via online training, as in \cite{shlezinger2019viterbinet}.
Furthermore, LoRD-Net differs from previously proposed deep unfolded MIMO receivers as surveyed in \cite{balatsoukas2019deep} in two key aspects: First, LoRD-Net is particularly designed for one-bit observations, being derived from the iterative optimization formulation which arises from such setups. Second, previous unfolded MIMO receivers either assumed prior knowledge of the channel parameters, as in \cite{samuel2019learning}, or alternatively, utilize external modules to directly estimate the CSI as in \cite{he2019model}. LoRD-Net exploits the fact that, for its unfolded relaxed convex optimization algorithm to yield the desired MLE, an alternative channel parameters, which differ from the true $\Theta$, should be  estimated. Consequently, the training procedure of LoRD-Net does not aim to recover the true CSI, but the one which yields a competitive objective which facilitates symbol detection, thus accounting for the overall system task. 

The proposed training procedure detailed in Algorithm~\ref{alg:Algo1} carries out each training stage once in a sequential manner. This strategy can be extended to optimizing the hyperparameters and the weights in an alternating fashion, i.e. repeating the stages multiple times, while using the learned $\bphi$ in Stage 2 in the Stage 1 that follows. Alternatively, the hyperparameters and the weights can be learned jointly in an end-to-end manner, by optimizing \eqref{eq:training2emprical} with respect to both $\Theta$ and $\bphi$ simultaneously. The main requirement for carrying out these training strategies compared to that detailed in Subsection \ref{subsec:train} is that the same number of layers $L$ should be used when learning both $\Theta$ and $\bphi$, while when these stages are carried out once sequentially, it is preferable to use large $L$ at Stage 1 and a smaller value, which dictates the number of learned weights, in Stage 2. Furthermore, our numerical evaluations show that training once in a two-stage fashion via Algorithm~\ref{alg:Algo1} yields similar and sometimes even improved performance compared to learning both $\Theta$ and $\bphi$ simultaneously in a one-stage manner, as well as when alternating between these two stages, as demonstrated in Section~\ref{sec:NumericalStudy}. 

A  possible extension of the training procedure is to account for ADCs with more than one bit, as well as allow LoRD-Net to optimize the quantization thresholds $\bb$ in light of the overall symbol recovery task. While accounting for multi-level ADCs is a rather simple extension achieved by reformulating the objective function \eqref{eq:MLObj}, optimizing the quantization thresholds requires modifying the overall training strategy. The challenge here is that modifying $\bb$ results in different one-bit measurements $\br$. In a communication setup, in which periodic pilots are transmitted, one can envision gradual optimization of $\bb$ between consecutive pilot sequences, using their corresponding one-bit observations to further optimize LoRD-Net.  The study of LoRD-Net with multi-level ADCs and optimized thresholds is left for future work.

\vspace{-0.2cm}
\section{Numerical Study}
\label{sec:NumericalStudy}
\vspace{-0.1cm}
In this section, we numerically evaluate LoRD-Net\footnote{The source code is available at: \url{https://github.com/skhobahi/LoRD-Net}.}, and compare its performance with state-of-the-art model-based and data-driven methodologies. As a motivating application for the proposed LoRD-Net, we focus on the evaluation of  LoRD-Net for blind symbol detection task in one-bit MIMO wireless communications. In the following, we first detail the considered one-bit MIMO simulation settings in Subsection~\ref{subsec:NumSetup}, after which we evaluate the receiver performance, compare LoRD-Net to alternative unfolded architectures, and numerically investigate its training procedure in Subsections \ref{subsec:NumRx}, \ref{subsec:NumUnfold}, and \ref{subsec:NumTrain}, respectively.  
.%, upon which we will evaluate the performance of UGD-Net in the context of one-bit MIMO communication in the following parts.

\vspace{-0.2cm}
\subsection{Simulation Setting}
\label{subsec:NumSetup}
\vspace{-0.1cm}
We consider an up-link one-bit multi-user MIMO scenario as in \eqref{eq:sysmodel}. We focus on a single cell in which a base station (BS) equipped with $m$ antenna elements serves  $n$ single-antenna users. Specifically, we consider two cases of $(m,n)=(128,16)$ and $(m,n)=(64, 10)$, i.e., a $128\times16$ and a $64\times 10$ MIMO channel setup.
 %The training of the UGD-Net deployed at the BS is assumed to be performed during the up-link transmission and using a training pilot sequence of length $B$ leading to a training data-set of the form $\{(\bx_p^i,\br_p^i)\}_{i=0}^{B-1}$, where $\bx_p^i$ denotes the vector of the transmitted symbols (known a priori at both the BS and UEs), and $\br_p^i$ represents the output of the channel after applying the one-bit quantization. In the rest of this paper, we interchangeably use \emph{pilot length} and \emph{training size} to refer to the sample size $B$. Moreover, 
 The transmitted symbols of the users, represented by the unknown vector $\bx$, are randomized in an independent and identically distributed (i.i.d.) fashion from a BPSK constellation set $\mathcal{M} = \{-1, +1\}$. % resulting in $\mathcal{M}^n = \{-1,+1\}^n$. Further note that for such a 
The projection mapping is thus $\mathcal{P}_{\mathcal{M}^n}(\bx) = \mathrm{sign}(\bx)$, where the $\mathrm{sign}$ function is applied element-wise on the vector argument.  In the sequel, we assume that while the channel matrix $\bH$, representing the CSI, is not available at the BS, the noise statistics $\bC$ are known and are fixed to $\bC = \bI$. Accordingly, our goal is to utilize LoRD-Net to recover the transmitted symbols from the one-bit measurements. Note that the proposed methodology can carry out the task of symbol detection even for the case in which the noise statistics $\bC$ is unknown.

\emph{Channel Models:}
We evaluate  LoRD-Net under two channel models: \emph{(i)} i.i.d. Rayleigh fading channels, where  $\bH\sim\mathcal{N}(\mathbf{0}, \bI)$; and \emph{(ii)} the COST-2100  massive MIMO channel \cite{flordelis2019massive}. The COST-2100 channel model is a realistic geometry-based stochastic  model which accounts for prominent characteristics of massive MIMO channels, and is considered to be an established benchmark for evaluating MIMO communication systems. We generate the channel matrices for the COST-2100 model for a narrow-band indoor scenario with closely-spaced users at $2.6$~GHz band, where the BS is equipped with a uniform linear array (ULA) that has $m$ omni-directional receive antenna elements.
The one-bit ADC operation uses zero thresholds,   i.e.  $\bb=\mathbf{0}$. We define the signal-to-noise ratio ($\mathrm{SNR}$) as:
\begin{align}
    \mathrm{SNR} = \mathds{E}\left\{\|\bH\bx\|_2^2\right\}/\mathds{E}\left\{\|\bn\|_2^2\right\}.
\end{align}
%The BS is assumed to be located at the center of a $20\mathrm{m}\times 20\mathrm{m}$ grid

\vspace{4pt}

\emph{Benchmark Algorithms:} As LoRD-Net combines both model-based and data-driven inference, we compare its performance with state-of-the-art model-based and data-driven methodologies in a one-bit MIMO receiver scenario. In particular, we use the following benchmarking detection algorithms:

\vspace{3pt}
\begin{itemize}
    \item The model-based nML proposed in \cite{choi2016near}. The nML algorithm is based on a convex relaxation of the conventional ML estimator, and requires the exact knowledge of the channel parameters $\Theta=\{\bH, \bC\}$. We set the number of iterations of the nML algorithm to $700$, and the step-size is chosen using a grid search method to further improve the performance of the nML, while the remaining parameters are those reported in \cite{choi2016near}.
    \item The data-driven Deep Soft Interference Cancellation (DeepSIC) methodology proposed in \cite{shlezinger2020deepsic}, with five learned interference cancellation iterations.  DeepSIC is  channel-model-agnostic and can be utilized for symbol detection in non-linear settings such as low-resolution quantization setups. Unlike LoRD-Net, which is designed particularly for observations of the form \eqref{eq:sysmodel} where $\Theta=\{\bH, \bC\}$ is unknown,  DeepSIC has no prior knowledge of neither the channel model nor its parameters.
\end{itemize}

%\noindent
%$\bullet$ 

%\noindent
%$\bullet$ %We use the architecture detailed in \cite{shlezinger2020deepsic} with $L=5$ interference cancellation iterations, resulting in a total of    $\mathcal{T}_{\mathrm{DeepSIC}}(n,m,L) = Ln[(m + (n-1))\times 60 + 60\times 30 + 30]$ trainable parameters. %Note that the total number of trainable parameters of the DeepSIC architecture grows \emph{quadratically} with respect to the number of users $n$.

\vspace{-1pt}

%\smallskip
\emph{LoRD-Net Setting:} The LoRD-Net receiver is implemented  with  $L=30$ layers. %For both training stages, we assume that the total number of layers/iterations of the UGD-Net is $L=30$. 
Recall that the first training stage of the LoRD-Net is concerned with finding a competitive objective by carrying out the training of the network over the unknown set of channel parameters  $\Theta=\{\bH,\bC\}$. Unless otherwise specified, we focus on the case where only $\bH$ is unknown, and the correlation matrix of the noise $\bC$ is available. %We stress again that the proposed UGD-Net can also handle unknown noise statistics as well (e.g., by setting $\Theta=\{\bH, \bC\}$).

 During the first training stage,  we set $\delta=0.01$, and recover $\Theta^\star$ based on the objective \eqref{eq:training1} using the Adam stochastic optimizer \cite{kingma2014adam} with a constant learning rate of $10^{-3}$. %Note that the preconditioning matrices employed in this training stage are facilitating finding a competitive objective function. 
 %Once the first training stage is completed, we fix the obtained set of parameters $\Theta$ and perform the training over the set of preconditioning matrices as follows. The second training stage is concerned with optimizing the set of positive semi-definite preconditioning matrices to accelerate the convergence of the UGD-Net to the optimal points (note that the architecture of the UGD-Net is derived based on a first-order optimization technique). In particular, we consider the learning of diagonal preconditioning matrices during the second training stage, i.e., we set the trainable parameters of the UGD-Net for this stage as $\bphi = \{\bG_i\}_{i=0}^{L-1}$, where $\bG_i = \bW_i^T\bW_i$, with $\bW_i = \mathrm{Diag}(\delta_i^0, \delta_i^1, \cdots, \delta_i^{n-1})$.
Next, we carry out the training of the LoRD-Net during the second stage according to the objective function defined in \eqref{eq:training2emprical} and over the set of trainable parameters $\bphi$, using the Adam optimizer with a learning rate of $10^{-4}$, and a mini-batch of size $512$. We consider the learning of diagonal pre-conditioning matrices (unfolded weights) during the second training stage. The network is trained for $400$ epochs during the first training stage, and $400$ epochs during the second training stage, with the same value of $L=30$  used in both stages. %Finally, we wish to make the remark that the training of the UGD-Net is carried out once, online, and based on the received pilot sequence at the BS, and upon the completion of the training procedure, the network can be used for inference purposes. Due to the importance of reproducible research, we have made all the codes implemented publicly available along with this paper.

\vspace{-0.2cm}
\subsection{Receiver Performance}
\label{subsec:NumRx}
\vspace{-0.1cm}
Here, we evaluate the performance of the proposed LoRD-Net, comparing it to the aforementioned benchmarks as well as examining its dependence on the number of training samples $B$. 
In particular, we numerically evaluate the bit-error-rate (BER) performance versus SNR using different training sizes $B\in\{1024, 2048\}$, for both $128\times 16$ and $64\times 10$ channel configurations. 
For DeepSIC, we use only $B=2048$, while the nML recever of \cite{choi2016near} operates with perfect CSI, i.e., with full accurate knowledge of $\Theta$.
All data-driven receivers are trained for each SNR  separately, using a dataset corresponding to that specific SNR value.

\begin{figure*}[!t]
	\centering
	\hspace{-.1cm}
	\subfigure[Rayleigh fading]{\includegraphics[width =0.42\linewidth]{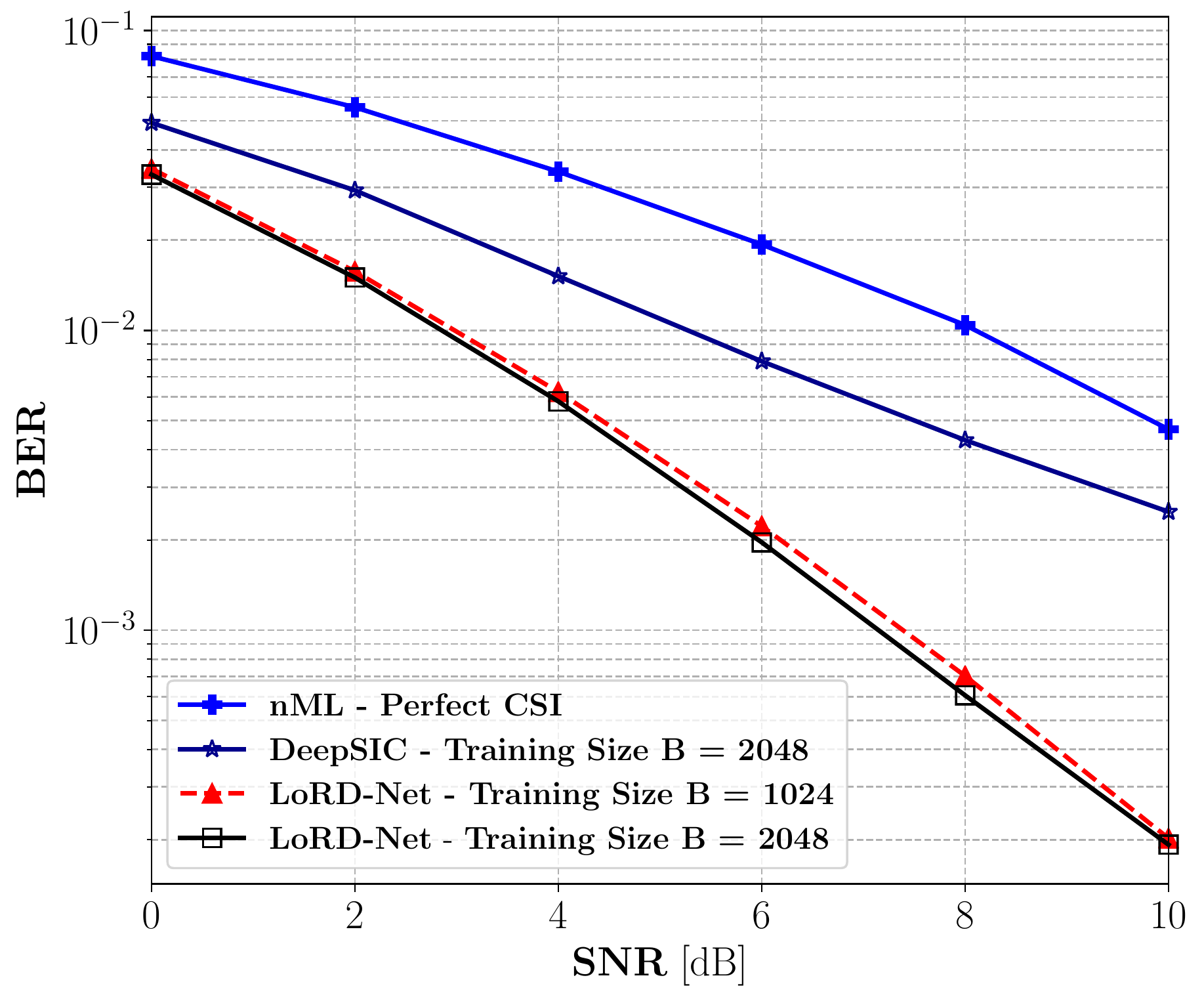}\label{fig:128x16BERvsSNRRayleigh}}
	\quad
	\subfigure[COST-2100 massive MIMO ]{\includegraphics[width =0.42\linewidth]{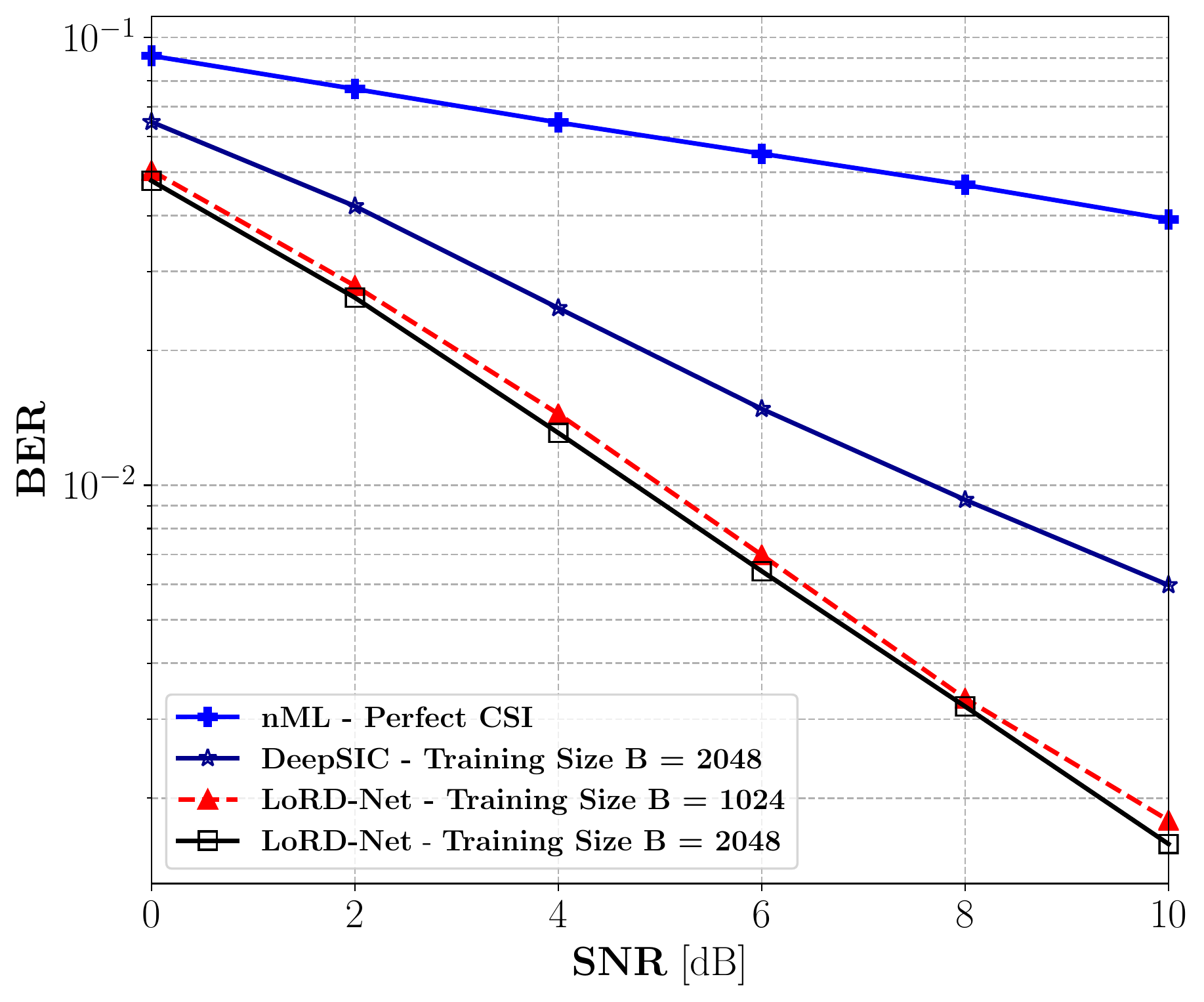}\label{fig:128x16BERvsSNRCOST2100}}
	\caption{BER performance versus SNR over a $128\times 16$ channel configuration.}
	\label{fig:128x16BERvsSNR}
	\vspace{-15pt}
\end{figure*}

\begin{figure*}[!t]
	\centering
	\hspace{-.1cm}
	\subfigure[Rayleigh fading]{\includegraphics[width =0.42\linewidth]{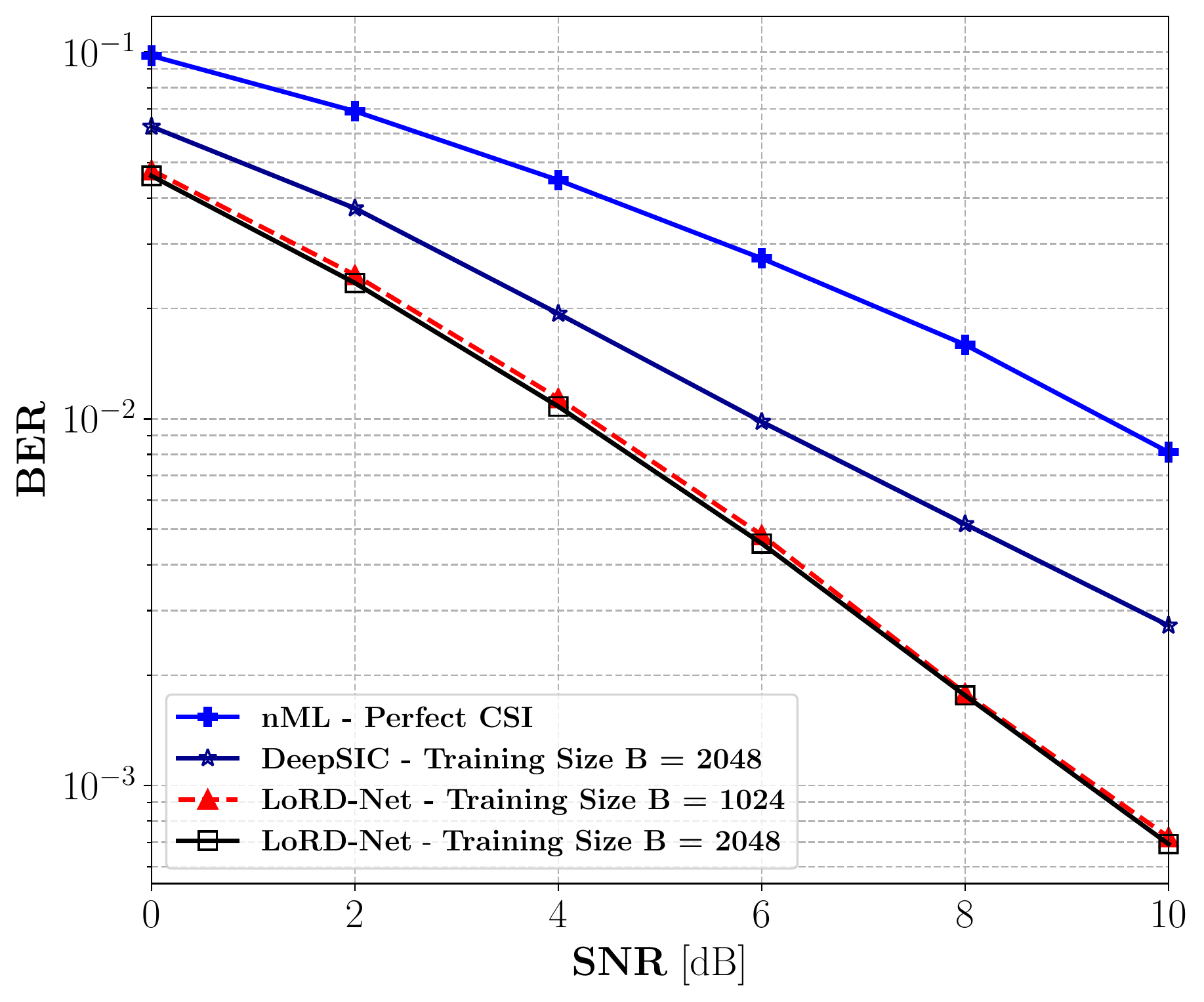}\label{fig:64x10BERvsSNRRayleigh}}
	\quad
	\subfigure[COST-2100 massive MIMO]{\includegraphics[width =0.42\linewidth]{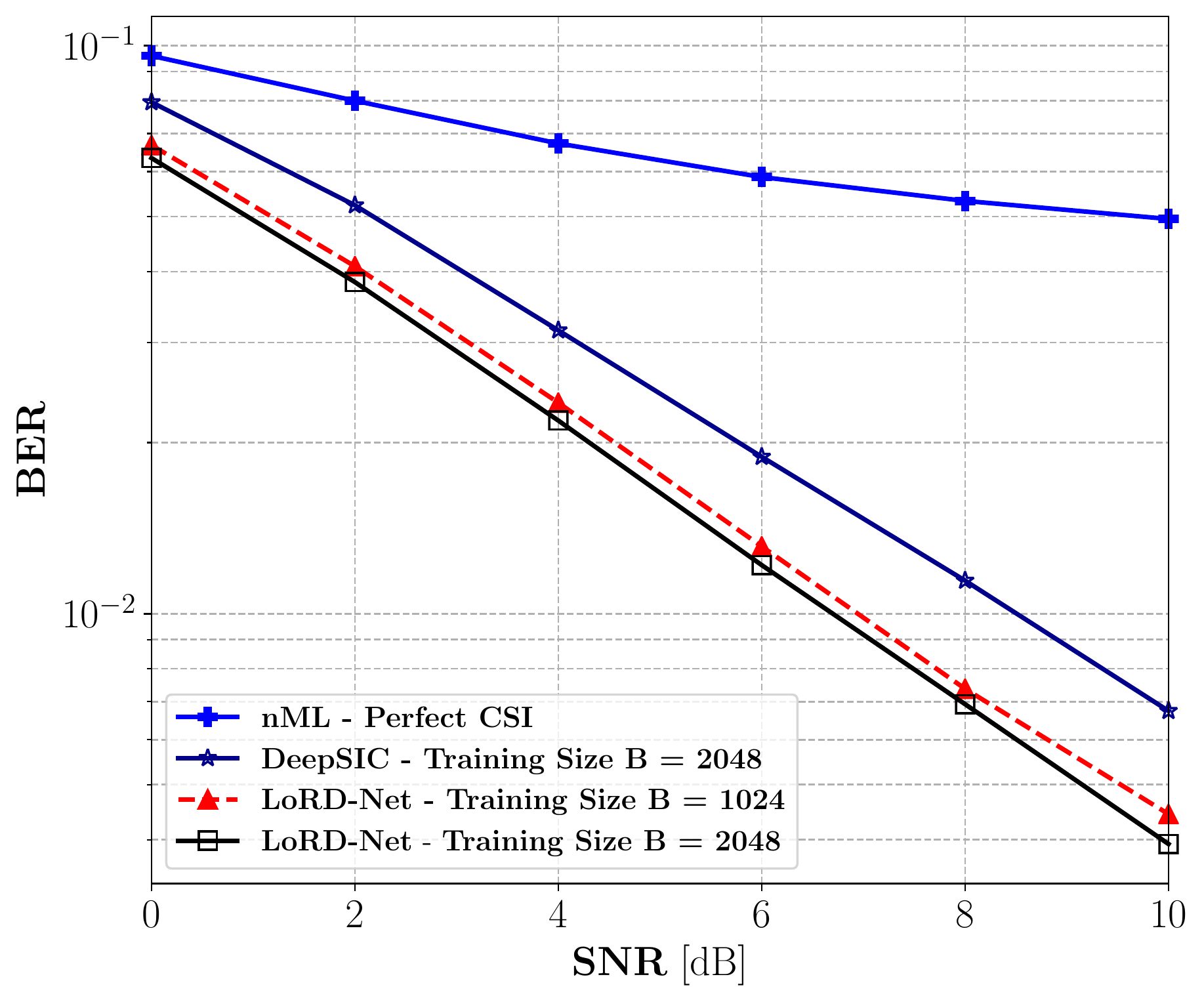}\label{fig:64x10BERvsSNRCOST2100}}
	\caption{BER performance versus SNR over a $64\times 10$ channel configuration.}
	\label{fig:64x10BERvsSNR}
	\vspace{-15pt}
\end{figure*}

The results are depicted in Figs.~\ref{fig:128x16BERvsSNRRayleigh} and \ref{fig:128x16BERvsSNRCOST2100} for a $128\times 16$ channel configuration under the Rayleigh fading  and COST-2100 channel models, respectively. Furthermore, the BER performance for a $64\times 10$ configuration under both channel models are illustrated in Fig. \ref{fig:64x10BERvsSNRRayleigh} for the Rayleigh fading channel, and in Fig. \ref{fig:64x10BERvsSNRCOST2100}, for the COST-2100 channel model.   
Based on the results presented in Figs.~\ref{fig:128x16BERvsSNR} and~\ref{fig:64x10BERvsSNR}, one can observe that LoRD-Net significantly outperforms the competing model-based and data-driven algorithms and achieves improved detection performance under both simulated channels, as well as both MIMO configurations.

In particular, the nML algorithm, which is designed to iteratively approach the MLE using ideal CSI (prior knowledge of the channel matrix), is notably outperformed by LoRD-Net. Such gains by LoRD-Net, which learns to compute the MLE from data without requiring CSI, compared to the model-based nML algorithm, demonstrate the benefits of learning a competitive objective function combined with a relaxed deep unfolded optimization process. % Namely, the nML algorithm employs the exact knowledge of the channel matrix and seeks to recover the true symbols by employing a convex relaxation of the underlying ML optimization problem (whose optimal points on the continuous domain might not coincide with that of the discrete set $\mathcal{M}^n$). On the other hand, the UGD-Net seeks to learn a competitive objective function (during the first training stage) by learning the unknown set of parameters $\Theta=\{\bH\}$ of the objective function based the received pilot sequence to bridge the gap between the underlying (non-convex) ML optimization problem (see Eq. \eqref{eq:originalopt}) and the relaxed one defined in \eqref{eq:optimization}, and further minimizes the gap between the solution of the two in continuous and discrete domain leading to a superior performance.
%At the first glance, the idea of not using the true channel matrix when dealing with relaxed optimization techniques for symbol recovery purposes might seem counter-intuitive. 
Specifically, the results depicted in Figs.~\ref{fig:128x16BERvsSNR}-\ref{fig:64x10BERvsSNR} illustrate that one can significantly improve the receiver performance by learning a new channel matrix $\bH$ upon which the learned competitive objective function admits optimal points near the true symbols. The learning of the competitive objective function  is possible due to the hybrid model-based/data-driven nature of  LoRD-Net, and the fact that it is derived based on the unfolding of first-order optimization techniques. % and by incorporating the relaxed optimization model into the architecture of the UGD-Net.
From a computational complexity point-of-view, the depicted performance of the nML algorithm in Figs. \ref{fig:128x16BERvsSNR}-\ref{fig:64x10BERvsSNR} is achieved by employing $700$ iterations of a first-order optimization algorithm, while LoRD-Net uses only $L=30$ layers/iterations---exhibiting a significant reduction in the computational cost during inference as compared to the nML algorithm. %Finally, note that the UGD-Net architecture significantly outperforms the nML algorithm even for pilot sequences as short as $B=256$.

Comparing LoRD-Net to DeepSIC illustrates that LoRD-Net benefits considerably from its model-aware architecture. The fact that LoRD-Net is particularly tailored to the one-bit system model of \eqref{eq:sysmodel} allows it to achieve improved accuracy, even in the case of training with small amounts of data. For instance, for the $128\times16$ MIMO Rayleigh fading channel (see Fig.~\ref{fig:128x16BERvsSNRRayleigh}), LoRD-Net trained with $B=2048$ samples, achieves BER of $10^{-2}$ at SNR of $3$dB, while DeepSIC trained with the same dataset requires SNR as high as $5$dB to achieve such an error rate. Considering Fig.~\ref{fig:128x16BERvsSNRCOST2100}, a similar behavior is observed in the COST-2100 channel, for a BER of $3 \times 10^{-2}$. A similar performance gain for LoRD-Net can be observed in a $64\times 10$ configuration; see  Fig.~\ref{fig:64x10BERvsSNR}. Furthermore, it can be observed that the LoRD-Net still outperforms the DeepSIC methodology, even when trained on $2$ times less training samples. %As mentioned earlier, this significant reduction in the training size of UGD-Net while maintaining performance gains stems from proper use of domain knowledge. %, which allows it to utilize a relatively small number of trainable parameters, which can in turn be accurately adapted using small training sets. 
In particular, for the $(128\times 16)$ channel setup considered in this part, the total number of trainable parameters of  LoRD-Net is merely ${|\Theta=\{\bH\}|} + {|\bphi|} = n\left(L+m\right) = 2528$. For comparison, DeepSIC, which uses and trains a multi-layer fully-connected network for each user at each interference cancellation iterations, consists here of over $8 \times 10^5$ trainable parameters.
Such a reduction in the number of parameters allows for achieving substantially improved performance with much smaller training points, as observed in Figs.~\ref{fig:128x16BERvsSNR}-\ref{fig:64x10BERvsSNR}. 
%Note that the UGD-Net for the case of $B=256$ under the COST-2100 channel model provided in Fig. \ref{fig:BERvsSNRCOST2100} achieves on-par performance compared to the DeepSIC architecture for $\mathrm{SNR}\in\{0,2,4\}$ and it outperforms the DeepSIC algorithm for higher SNRs, with $8$ times less number of training samples and with $300$ times less number of trainable parameters. Furthermore, the UGD-Net trained on pilot sequences of size $B\in\{512, 1024, 2048\}$ achieves a substantially better performance than that of both the data-driven DeepSIC and model-based nML algorithm as it can be clearly observed from Fig. \ref{fig:BERvsSNR}. 
Finally, we note that the small number of trainable parameters of LoRD-Net % which result in its ability to train accurately with small data sets, 
shows its potential for online or real-time training, as proposed in \cite{shlezinger2019viterbinet}. This can be achieved by using periodic pilots with minimal overhead on the communication, while inducing a relatively low computational burden in its periodic retraining. %\nocite{soltanalian2016training,hassibi2003much}  %  allows the network can be trained with much less number of optimization epochs (e.g., the UGD-Net is trained for $1200$ epochs) as compared to the conventional black-box DNN models and the DeepSIC architecture---making the UGD-Net a great potential candidate for real-time signal processing applications.

\begin{figure}
	\centering
	\includegraphics[width =\figwidth]{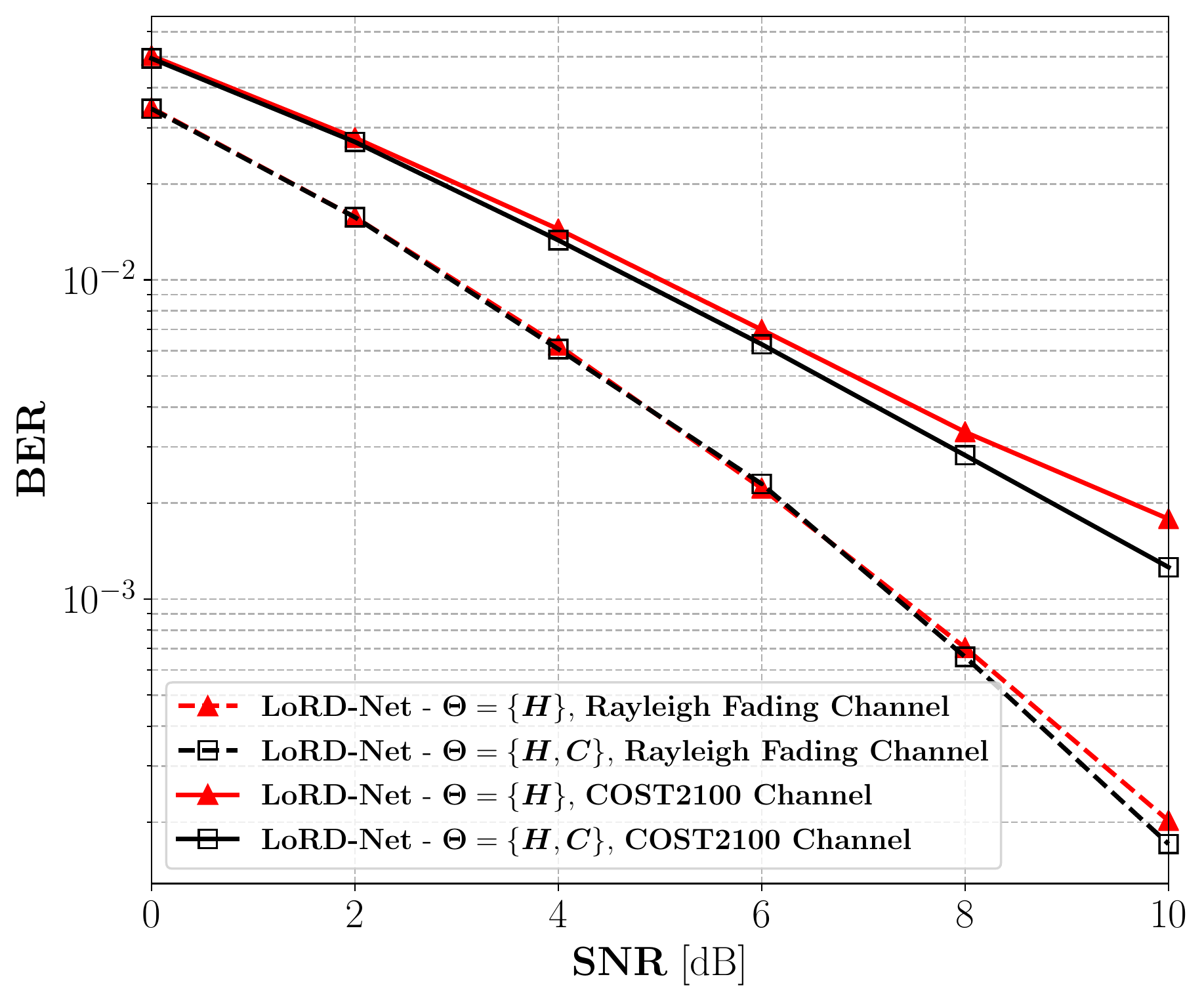}
	\caption{BER versus $\mathrm{SNR}$ for both channel models and a training size of $B=1024$. The performance of the proposed LoRD-Net is provided for both scenarios of training over $\Theta=\{\bH\}$ (i.e., known noise statistics $\bC$), and over $\Theta=\{\bH,\bC\}$ corresponding to unknown channel matrix and noise statistics.} 
	\label{fig:BERvsSNRHC}
	%\vspace{10pt}
\end{figure}

So far, we have investigated the performance of the proposed LoRD-Net for scenarios with known noise statistics, and unknown $\bH$ (i.e., $\Theta=\{\bH\}$). Next, we investigate the detection performance of LoRD-Net when both the channel and noise covariance matrices are not available, i.e., we set $\Theta=\{\bH,\bC\}$ and carry out the training according to the proposed two stage methodology. Specifically, we consider the learning of a diagonally structured $\bC$ in addition to the channel matrix $\bH$ for this scenario. Fig.~\ref{fig:BERvsSNRHC} demonstrates the BER versus $\mathrm{SNR}$ performance of  LoRD-Net  under both channel models, when trained using  a dataset of size $B=1024$. The performance of LoRD-Net for the case of $\Theta=\{\bH\}$ is further provided for comparison purposes. Observing Fig.~\ref{fig:BERvsSNRHC}, one can readily conclude that the proposed network can successfully perform the task of symbol detection also when $\bC$ in unknown. Furthermore, it can be observed that a small gain in performance is achieved for both channel models when $\Theta=\{\bH,\bC\}$ as compared to the case of $\Theta=\{\bH\}$, which is presumably due to the careful addition of more degrees of freedom in learning a competitive surrogate model.

\vspace{-0.2cm}
\subsection{Performance of Competing Deep Unfolded Architectures}
\label{subsec:NumUnfold}
\vspace{-0.1cm}
In this part, we compare the performance of the proposed LoRD-Net with alternative deep unfolding-based architectures tailored for the problem at hand. Recall that the architecture of LoRD-Net uses trainable parameters which are shared among the different layers, as illustrated in Fig.~\ref{fig:UGD_Net1}. Thus, LoRD-Net is comprised of a relatively small number of trainable parameters, and uses a two-stage learning method to train the shared parameters, representing the competitive model, and the iteration-specific weights, encapsulating the first-order optimization coefficient. Nonetheless, the conventional approach for unfolding first-order optimization techniques is to over-parameterize the iterations, and then, train in an end-to-end manner using a one-stage training procedure discussed earlier. Therefore, to numerically evaluate the proposed unfolding mechanism of LoRD-Net, we next compare it to two conventional unfolding based benchmarks derived from the relaxed MLE:
\begin{itemize}
    \item {\em Benchmark 1}: An over-parameterized deep unfolded architecture obtained by setting the computational dynamics for the $i$th layer as:
\begin{align} %\label{eq:objective_grad}
    \bar{g}_{\phi_i}(\bx_i;\br)= \bx_i - \bA_i\bR\;\boldsymbol\eta \left(\bR\left(\bb - \bB_i\bx_i\right) \right)\label{eq:GDUD}.
\end{align}
    Here, $\phi_i=\{\bA_i\in\mathds{R}^{n\times m}, \bB_i\in\mathds{R}^{m\times n}\}$ are the trainable parameters of the $i$th layer, and $\bR=\mathrm{Diag}(\br)$.
    \item {\em Benchmark 2}: Here, we again use the unfolded architecture given in \eqref{eq:GDUD}, while limiting the number of trainable parameters by constraining the rank of the learned matrices. In particular, we set $\bA_i = \bP_i\bQ_i$ and $\bB_i=\bR_i\bS_i$, where $\phi_i = \{\bP_i\in\mathds{R}^{n\times r}, \bQ_i\in\mathds{R}^{r\times m}, \bR_i\in\mathds{R}^{m\times r}, \bS_i\in\mathds{R}^{r\times n}\}$ denotes the set of trainable parameters of the $i$th layer of the unfolded network. The dimension $r < \mathrm{min}(m,n)$ controls the rank of the resulting weight matrices $\{\bA_i, \bB_i\}$, and thus the number of trainable parameters.
\end{itemize}

Comparing  \eqref{eq:GDUD} with the corresponding dynamics of LoRD-Net in \eqref{eq:proposed1}, we note that the channel matrix $\bH$, the pre-conditioning matrices $\bG_i$, and the noise covariance matrix $\bC$ are now absorbed into the \emph{per-layer} trainable matrices $\bA_i$ and $\bB_i$. Accordingly, these unfolded benchmarks, which follow the conventional approach for unfolding optimization algorithms, are less faithful to the underlying model. 
These benchmarks also differ from LoRD-Net in their number of trainable parameters. In particular, Benchmark 1 with $L$ layers has $2Lnm$ trainable parameters, while Benchmark 2 has  $2Lr(m+n)$ weights, which can be controlled by the setting of the hyperparameter $r$. For comparison,   LoRD-Net has $n(L+m)$ trainable parameters for the case of $\Theta=\{\bH\}$ and diagonally structured pre-conditioning matrices, while for the case of $\Theta= \{\bH,\bC\}$ with a diagonally structured pre-conditioning matrix and noise covariance matrix it has  $n(L+m) +m$ trainable parameters.

We evaluate the performance of the proposed LoRD-Net compared to the unfolded benchmarks in the following simulation setup. We consider train all the considered network using a dataset of size $B=1024$, while the highly-parameterized Benchmark 1 is also trained using $B=2048$ samples. For Benchmark 2, we set $r=1$. All architectures  have $L=30$ layers and their performance are evaluated on the same testing dataset of size $B=2048$. The unfolded benchmarks are trained in the conventional end-to-end fashion. The channel model is a $(128\times 16)$ Rayleigh fading channel. Foror the considered scenario above, the LoRD-Net admits a total of $2528$ trainable parameters, while Benchmark 1 ha a total of $122880$ (approximately $50$ times more parameters than LoRD-Net), while Benchmark 2 has $8640$ trainable parameters.

\begin{figure}
	\centering
	\includegraphics[width =\figwidth]{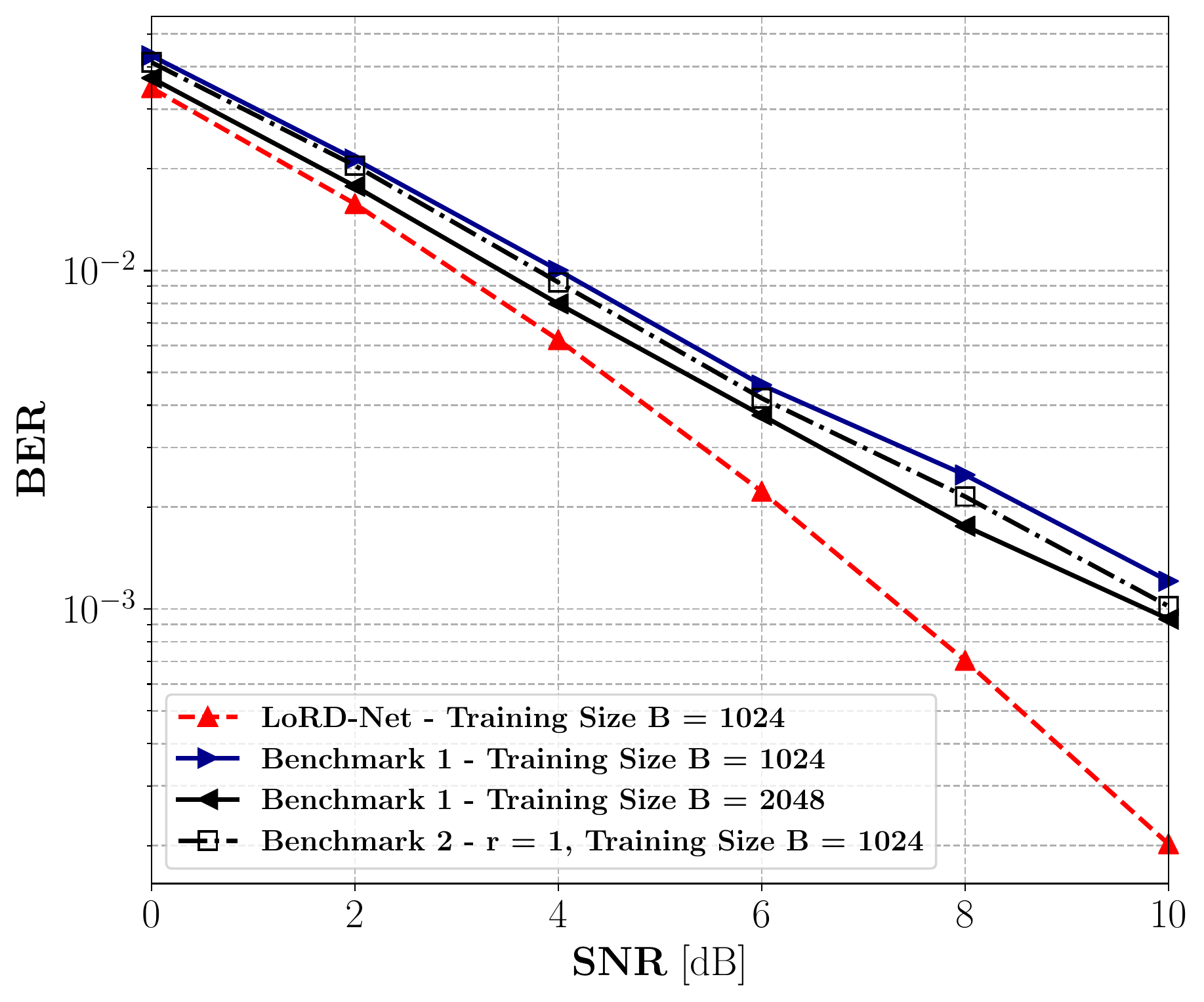}
	\caption{BER versus $\mathrm{SNR}$ of LoRD-Net compared to the unfolded benchmark for a $(128\times 16)$  Rayleigh fading channel model.} 
	\label{fig:GDUD-Net}
	%\vspace{10pt}
\end{figure}

Fig~\ref{fig:GDUD-Net} depicts the BER versus $\mathrm S\mathrm N\mathrm R$ of LoRD-Net compared to the unfolded benchmarks.  We observe in Fig.~\ref{fig:GDUD-Net} that the proposed LoRD-Net significantly outperforms the conventional unfolding based benchmarks,indicating the gains of the increased level of domain knowledge Incorporated in to the architecture of LoRD-Net and its two stage training procedure. It is also observed that the performance of Benchmark 1 increases with more training samples. Interestingly, for a small training set of $B=1024$ samples, Benchmark 2, which is obtained by imposing a rank constraint on the trainable parameters of Benchmark 1, achieves improved performance over Benchmark 1, due to its notable reduction in the number of trainable parameters. 

\color{black}

%\vspace{-1pt}
\vspace{-0.2cm}
\subsection{Training Analysis}
\label{subsec:NumTrain}
\vspace{-0.1cm}
In this part, we analyze the performance of the proposed two-stage training procedure described in Subsection \ref{subsec:train}. The training aspects of LoRD-Net are numerically evaluated for the $128\times 16$ Rayleigh channel model detailed before. % and investigate the how the performance of UGD-Net scales with the total number of training samples $B$. Throughout this subsection and similar to the previous numerical experiments, we consider a $128\times 16$ one-bit MIMO system with BPSK modulation under the Rayleigh fading channel model assumption, and assume that the channel matrix is unknown and we set $\Theta=\{\bH\}$ during the first training stage of the UGD-Net.

\begin{figure}
	\centering
	\includegraphics[width =\figwidth]{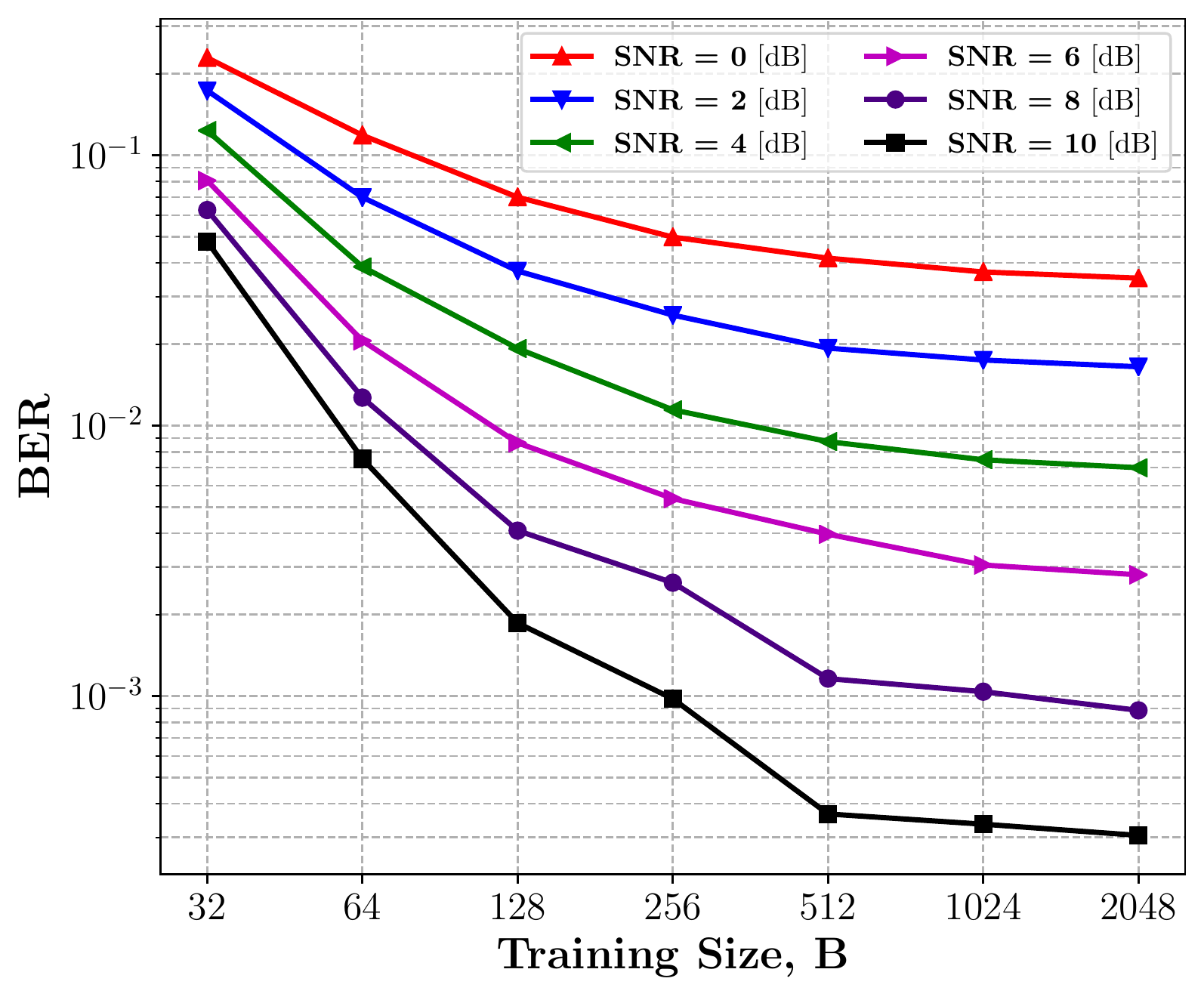}
	\caption{BER versus training size $B$ for the Rayleigh fading channel.}
	\label{fig:BERvsB}
	%\vspace{10pt}
\end{figure}

Following our insight on the ability of LoRD-Net to accurately train with small datasets, we begin by evaluating the performance of the LoRD-Net versus the training data size~$B$. For this study, we generate training datasets of size $B\in\{32, 64, 128, 256, 512, 1024, 2048\}$ and evaluate the performance of  LoRD-Net using $2048$ test samples. Fig.~\ref{fig:BERvsB} depicts the BER achieved for each training size $B$,  for $\mathrm{SNR} \in \{0,2,4,6,8,10\}$~dB.  
%We first begin by investigating the performance of the UGD-Net versus the training size $B$. For this study, we create training datasets of size $B\in\{32, 64, 128, 256, 512, 1024, 2048\}$ and evaluate the performance of the UGD-Net on testing datasets of size $2048$, for $\mathrm{SNR} \in \{0,2,4,6,8,10\}$. 
%Fig. \ref{fig:BERvsB} illustrates the BER against the training size $B$ for various SNR values. 
We can observe from Fig.~\ref{fig:BERvsB}  that the  performance of the LoRD-Net improves across all $\mathrm{SNR}$ values, where the improvements are most notable for $B\leq 256$. Interestingly, it may be concluded from Fig.~\ref{fig:BERvsB} that LoRD-Net is capable of accurately and reliably performing the task of symbol detection without CSI with as few as $B=512$ samples. The ability of LoRD-Net to train with very few training samples (compared to the black-box DNN models for one-bit MIMO receivers\cite{balevi2019one, zhang2020deep}, as well as  the DeepSIC architecture), stems from its incorporation of the domain-knowledge in designing the LoRD-Net architecture. This in turn leads to far fewer trainable parameters requiring much less training samples for optimizing the network.

Next, we analyze the performance and the effect of the two stage training methodology detailed in Algorithm~\ref{alg:Algo1} on the detection performance of the LoRD-Net architecture. %In particular, we seek to shed some light on the rationale behind the adopted training methodology.
Recall that the first training stage is concerned with finding a competitive objective function through an optimization of LoRD-Net over the unknown system parameters $\Theta$, while the second training stage tunes the positive definite preconditioning matrices $\bphi=\{\bG_i\}$  to accelerate the convergence of the LoRD-Net to the optimal points of the obtained competitive objective. 
To numerically evaluate the performance of the training methodology, we set $\mathrm{SNR}=8$ dB, and generate a training dataset of size $B=512$ and a testing dataset of size $2048$. Then, we compare performance of Algorithm~\ref{alg:Algo1} with two other competing training procedures:

\noindent
$\bullet$ \emph{One-Stage Training}: Here, the weights $\bphi$ and the unknown system parameters $\Theta$ are jointly learned in a single stage. The objective of this one stage training procedure for LoRD-Net is
\begin{align}\label{eq:training2emprical2}
    \underset{\bphi=\{\bG_l\}_{l}, \Theta\in\bTheta}{\mathrm{min}}\;\;\frac{1}{B}\sum_{i=0}^{B-1}\left\|\mathcal{G}^{L}_{\bphi}(\bx_0;\Theta,\br_p^i) - \bx_p^i\right\|_2^2.
\end{align}
%We note that the above training procedure corresponds to jointly learning a competitive objective function and the optimization policy $\bphi$. From a statistical learning perspective, the above model has more training variables in a single stage of training as opposed to the proposed two stage training procedure with each stage having less parameters than the total number of free variables in the entire model. Accordingly, in a limited data regime, once the training variables increases (and so the capacity of the network), the above model is prone to overfitting, i.e., admitting lower training error with higher testing error as opposed to other models.
$\bullet$ \emph{Alternating Training}:
This procedure is concerned with training the network by alternating between the two optimization problems \eqref{eq:training1} and \eqref{eq:training2emprical} consecutively with respect to each training epoch. Here, the network is trained over $400$ alternations, corresponding to a total of $800$ training epochs. Namely, at each epoch index $i$, we update the variables $\Theta$  for odd $i$ and update $\bphi$ for even $i$.

%Fig. \ref{fig:BERvsTraining} depicts the BER versus the training epoch for both the training and testing dataset. We first note that the proposed two-stage training method (Algorithm~\ref{alg:Algo1}) outperforms the competing procedures, yielding lower testing error values. Interestingly, we observe that the proposed methodology successfully closes the generalization gap as the testing and training error are very close to each other. On the other hand, the other two training procedures admit relatively large generalization gaps, indicating the fact that their utilization has resulted in an over-fitting of the network to the data. Furthermore, it can be observed from Fig. \ref{fig:BERvsTraining} that the major improvement of the detection accuracy of  LoRD-Net is taking place during the first training stage when finding a competitive objective function, where a slight improvement in the BER is achieved during the second training stage, starting at epoch index $80 (\times 5)$.

\begin{figure}
	\centering
	\includegraphics[width =\figwidth]{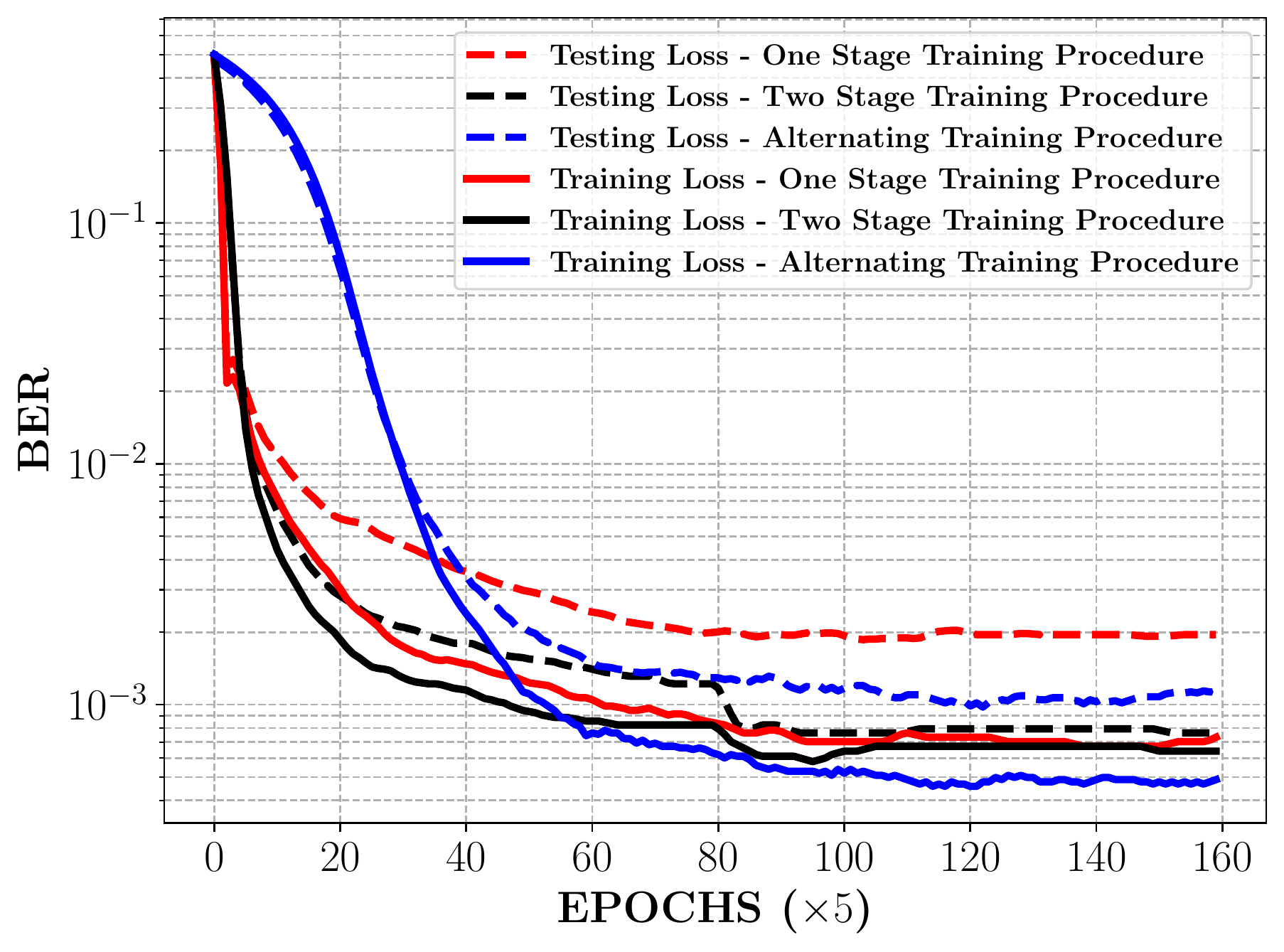}
	\caption{BER versus the training epoch number of  LoRD-Net, Rayleigh fading channel,  ${\rm SNR}=8$ dB.}
	\label{fig:BERvsTraining}
\end{figure}

Before we proceed with the evaluation results, we provide some useful connections to notions widely used in the deep learning literature. Generally speaking, the performance of a statistical learning framework and its training procedure is evaluated using its generalization gap and testing error. The \emph{generalization gap} of a model can be defined as the difference between the training and testing errors. Specifically, a model with smaller generalization gap and smaller testing error is highly favourable. Furthermore, a higher generalization gap may indicate that the network has over-fitted to the data, and hence, it does not generalize well. For two models with the same generalization gap, the one with lower testing error is favourable. Fig. \ref{fig:BERvsTraining} depicts the BER versus the training epoch for both the training and testing dataset. We first note that the proposed two stage training method outperforms all other competing procedures and it assumes a significantly lower testing error as compared to other algorithms. Interestingly, one can observe that the proposed methodology has successfully closed the generalization gap as the testing and training error are very close to each other. On the other hand, the other two training procedures admits very large generalization gap indicating the fact that their utilization has resulted in an over-fitting of the network to the data. Furthermore, it can be observed from Fig. \ref{fig:BERvsTraining} that the major improvement of the detection accuracy of the LoRD-Net is taking place during the first training stage when finding a competitive objective function, i.e., epochs $i < 80 (\times 5)$, where a slight improvement in the BER is achieved during the second stage, i.e., $i\geq80 (\times 5)$.

The success of the proposed two stage training procedure in closing the generalization gap compared to the one stage training procedure is presumably due to the fact that the two-stage training approach leads to an \emph{implicit regularization} on the model capacity limiting the total number of parameters used during the entire training procedure. On the contrary, the one stage training procedure allows the neural network to use its full capacity leading to an over-fitting and a larger generalization gap, as observed in Fig. \ref{fig:BERvsTraining}.

% In order to understand the behaviour and the information gain of the UGD-Net during the second training procedure, we need to dive deeper into a per-layer BER evalutation and transition of the performance of the UGD-Net during both training stages---see below for more details.

\begin{figure*}
	\centering
	%\hspace{-.1cm}
	\subfigure[Rayleigh fading]{\includegraphics[width =0.42\linewidth]{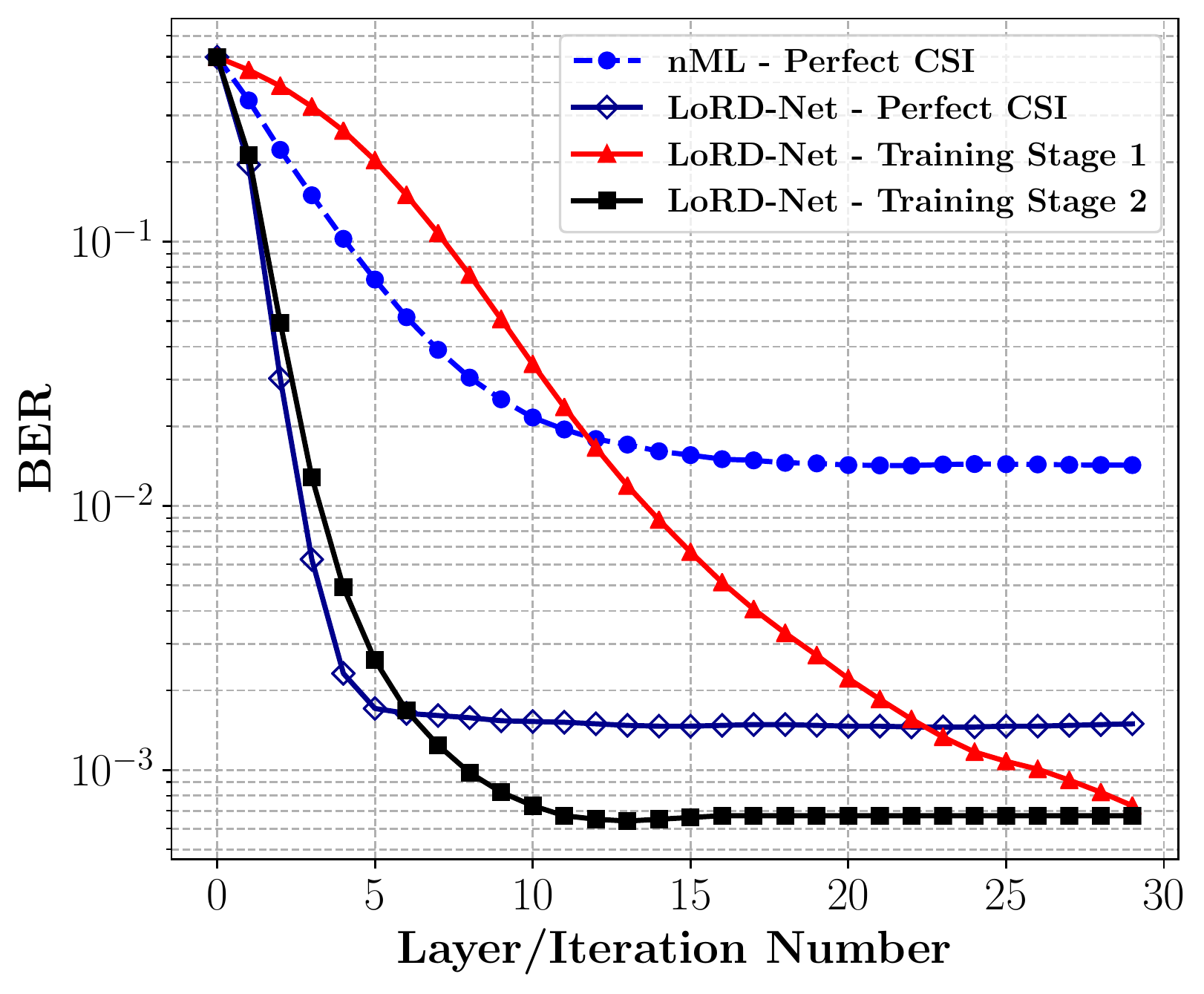}\label{fig:BERvsLayerRayleigh}}
	\qquad\quad
		\subfigure[COST-2100 Massive MIMO]{\includegraphics[width =0.42\linewidth]{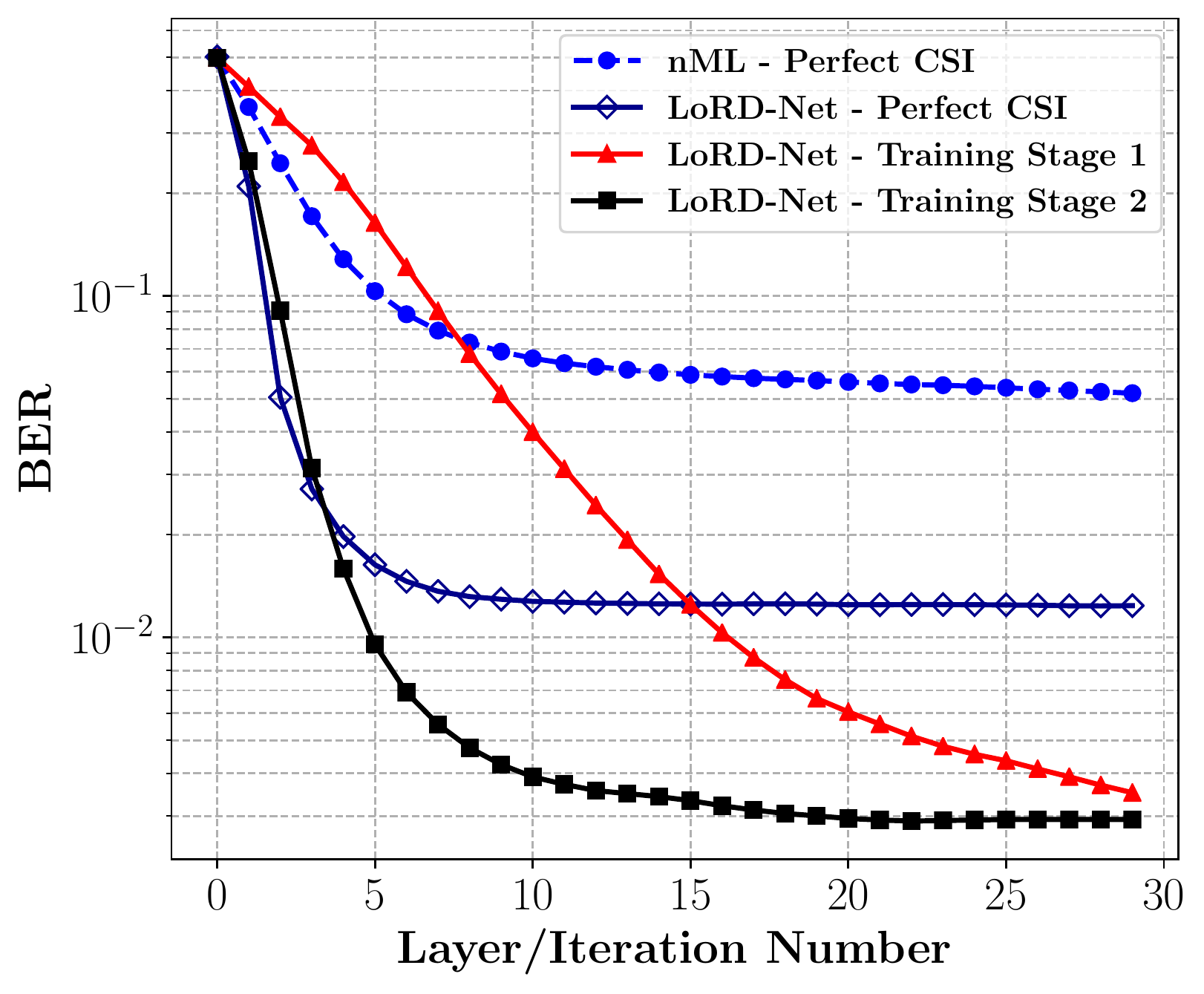}\label{fig:BERvsLayerCOST2100}}
	\caption{BER performance of LoRD-Net after completing training stages 1 and 2 versus the layer/iteration number for (a) the Rayleigh fading channel,  and (b) the COST-2100 massive MIMO channel, with ${\rm SNR}=8$ dB.}
	\label{fig:BERvsLayer}
	\vspace{-0.3cm}
\end{figure*}

As discussed in Subsection \ref{subsec:train}, the second training stage allows LoRD-Net to achieve fast inference, i.e., accelerated  convergence to the optimal points of the competitive objective function. To illustrate this behavior, we perform a per-layer BER evaluation of LoRD-Net, exploiting the interpretable model-based nature of the LoRD-Net, in which each layer represents an unfolded first-order optimization iteration, and thus its output can be used as an estimate of the transmitted symbols. %Note that such an analysis is made possible due to the model-based nature of the UGD-Net and the fact that it is an interpretable deep architecture (unlike its black-box DNN counterparts). In particular, 
Figs.~\ref{fig:BERvsLayerRayleigh} and \ref{fig:BERvsLayerCOST2100} depict the BER versus the layer/iteration number of LoRD-Net at the completion of training stages 1 and 2, for the Rayleigh fading channel and the COST-2100 channel model, respectively. We observe in Fig. \ref{fig:BERvsLayer} that the convergence of LoRD-Net  after the completion of the first training stage is slow and requires at least $L=30$ layers/iterations to converge. Interestingly,  we note from Fig. \ref{fig:BERvsLayer} that the second training stage indeed results in an acceleration of the convergence of  LoRD-Net via learning the best set of pre-conditioning matrices for the problem at hand in an end-to-end manner. In particular, after the completion of the second training stage, LoRD-Net can accurately and reliably recover the symbols with as few as $10$ layers. This observation hints that one can consider further truncation of the LoRD-Net after the training to reduce the computational complexity while maintaining its superior performance. 

In order to quantify the quality of the learned competitive objective in closing the gap between the discrete optimization problem and its continuous version, we further provide the per-iteration performance of the nML algorithm and the LoRD-Net algorithm which operate with perfect CSI. For this scenario, LoRD-Net utilizes the true $\Theta$, and is thus optimizer only over the weights $\bphi$ while employing the exact channel model $\bH$. It is observed from Figs.~\ref{fig:BERvsLayerRayleigh}-\ref{fig:BERvsLayerCOST2100} that learning a new surrogate model for the continuous optimization problem at hand is indeed highly beneficial and admits a far superior performance in recovering the transmitted symbols. % as compared to realizing the objective function $\calF_{\Theta}$ on the true system parameters $\Theta^{\star}$.%In other words, in applications where the computational power is limited, one can consider feed-forwarding the information through the UGD-Net for only $l\ll L$ layers. 
The analysis provided in Fig.~\ref{fig:BERvsLayer} further supports the rationale behind the proposed two-stage training methodology, and the fact that the second training stage results in an acceleration of the underlying first-order optimization solver (i.e., achieving a much faster descent per step) upon which the layers of the LoRD-Net are based.
% \begin{figure}[!t]
% 	\centering
% 	\includegraphics[width =\figwidth]{figs/UGD_Net_Final/Final/AllInOne.pdf}
% 	\caption{BER versus the training epoch number of  UGD-Net, Rayleigh fading channel,  ${\rm SNR}=8$ dB.}
% 	\label{fig:BERvsTraining}
% \end{figure}

%Indeed, the second training stages ensures that the output of the $l$-th layer of the UGD-Net is the most accurate estimation of the true transmitted symbols achievable by performing a first-order optimization technique upon which the layers of the UGD-Net is designed.

\vspace{-0.2cm}
\section{Conclusion}
\label{sec:Conclusion}
\vspace{-0.1cm}
In this work, we introduced  LoRD-Net, which is a hybrid data-driven and model-based deep architecture for blind symbol detection from one-bit observations. The proposed methodology is based the unfolding of  first-order optimization iterations for the recovery of the MLE. We proposed a two-stage training procedure incorporating the learning of a competitive objective function, for which the unfolded network yields an accurate recovery of the transmitted symbols from one-bit noisy measurements. In particular, owing to its model-based nature, LoRD-Net has far fewer trainable parameters compared to its data-driven counterparts, and can be trained with very few training samples. Our numerical results demonstrate that the proposed LoRD-Net architecture outperforms the state-of-the-art model-based and data-driven symbol detectors in multi-user one-bit MIMO systems. 
We also numerically illustrate the benefits of the proposed two-stage training procedure, which allows to train with small training sets and infer quickly, due to its interpretable model-aware nature.

\bibliographystyle{IEEEtran}
\bibliography{IEEEabrv,refs,refs_journal}

\begin{thebibliography}{10}
\providecommand{\url}[1]{#1}
\csname url@samestyle\endcsname
\providecommand{\newblock}{\relax}
\providecommand{\bibinfo}[2]{#2}
\providecommand{\BIBentrySTDinterwordspacing}{\spaceskip=0pt\relax}
\providecommand{\BIBentryALTinterwordstretchfactor}{4}
\providecommand{\BIBentryALTinterwordspacing}{\spaceskip=\fontdimen2\font plus
\BIBentryALTinterwordstretchfactor\fontdimen3\font minus
  \fontdimen4\font\relax}
\providecommand{\BIBforeignlanguage}[2]{{%
\expandafter\ifx\csname l@#1\endcsname\relax
\typeout{** WARNING: IEEEtran.bst: No hyphenation pattern has been}%
\typeout{** loaded for the language `#1'. Using the pattern for}%
\typeout{** the default language instead.}%
\else
\language=\csname l@#1\endcsname
\fi
#2}}
\providecommand{\BIBdecl}{\relax}
\BIBdecl

\bibitem{8683876}
S.~{Khobahi}, N.~{Naimipour}, M.~{Soltanalian}, and Y.~C. {Eldar}, ``Deep
  signal recovery with one-bit quantization,'' in \emph{Proc. IEEE ICASSP}, May
  2019, pp. 2987--2991.

\bibitem{eldar2015sampling}
Y.~C. Eldar, \emph{Sampling theory: Beyond bandlimited systems}.\hskip 1em plus
  0.5em minus 0.4em\relax Cambridge University Press, 2015.

\bibitem{walden1999analog}
R.~H. Walden, ``Analog-to-digital converter survey and analysis,'' \emph{{IEEE}
  J. Sel. Areas Commun.}, vol.~17, no.~4, pp. 539--550, 1999.

\bibitem{andrews2014will}
J.~G. Andrews, S.~Buzzi, W.~Choi, S.~V. Hanly, A.~Lozano, A.~C. Soong, and
  J.~C. Zhang, ``What will 5{G} be?'' \emph{{IEEE} J. Sel. Areas Commun.},
  vol.~32, no.~6, pp. 1065--1082, June 2014.

\bibitem{jeon2018one}
Y.-S. Jeon, N.~Lee, S.-N. Hong, and R.~W. Heath, ``One-bit sphere decoding for
  uplink massive {MIMO} systems with one-bit {ADCs},'' \emph{{IEEE} Trans.
  Wireless Commun.}, vol.~17, no.~7, pp. 4509--4521, 2018.

\bibitem{rao2020massive}
S.~Rao, G.~Seco-Granados, H.~Pirzadeh, and A.~L. Swindlehurst, ``Massive {MIMO}
  channel estimation with low-resolution spatial sigma-delta {ADCs},''
  \emph{arXiv preprint arXiv:2005.07752}, 2020.

\bibitem{ameri2019one}
A.~Ameri, A.~Bose, J.~Li, and M.~Soltanalian, ``One-bit radar processing with
  time-varying sampling thresholds,'' \emph{{IEEE} Trans. Signal Process.},
  vol.~67, no.~20, pp. 5297--5308, 2019.

\bibitem{jin2020one}
B.~Jin, J.~Zhu, Q.~Wu, Y.~Zhang, and Z.~Xu, ``One-bit {LFMCW} radar: spectrum
  analysis and target detection,'' \emph{{IEEE} Trans. Aerosp. Electron.
  Syst.}, 2020.

\bibitem{xi2020bilimo}
F.~Xi, N.~Shlezinger, and Y.~C. Eldar, ``{BiLiMO}: Bit-limited {MIMO} radar via
  task-based quantization,'' \emph{arXiv preprint arXiv:2010.00195}, 2020.

\bibitem{xiao2019one}
P.~Xiao, B.~Liao, and J.~Li, ``One-bit compressive sensing via {S}chur-concave
  function minimization,'' \emph{{IEEE} Trans. Signal Process.}, vol.~67,
  no.~16, pp. 4139--4151, 2019.

\bibitem{khobahi2019model}
S.~Khobahi and M.~Soltanalian, ``Model-based deep learning for one-bit
  compressive sensing,'' \emph{{IEEE} Trans. Signal Process.}, vol.~68, pp.
  5292--5307, 2020.

\bibitem{shlezinger2018hardware}
N.~Shlezinger, Y.~C. Eldar, and M.~R. Rodrigues, ``Hardware-limited task-based
  quantization,'' \emph{{IEEE} Trans. Signal Process.}, vol.~67, no.~20, pp.
  5223--5238, 2019.

\bibitem{salamatian2019task}
S.~Salamatian, N.~Shlezinger, Y.~C. Eldar, and M.~M{\'e}dard, ``Task-based
  quantization for recovering quadratic functions using principal inertia
  components,'' in \emph{Proc. IEEE ISIT}, 2019.

\bibitem{shlezinger2019deep}
N.~Shlezinger and Y.~C. Eldar, ``Deep task-based quantization,''
  \emph{Entropy}, vol.~23, no.~1, 2021.

\bibitem{shlezinger2020learning}
N.~Shlezinger, R.~J.~G. van Sloun, I.~A.~M. Hujiben, G.~Tsintsadze, and Y.~C.
  Eldar, ``Learning task-based analog-to-digital conversion for {MIMO}
  receivers,'' in \emph{Proc. IEEE ICASSP}, 2020.

\bibitem{neuhaus2020task}
P.~Neuhaus, N.~Shlezinger, M.~D{\"o}rpinghaus, Y.~C. Eldar, and G.~Fettweis,
  ``Task-based analog-to-digital converters,'' \emph{arXiv preprint
  arXiv:2009.14088}, 2020.

\bibitem{gong2019rf}
T.~Gong, N.~Shlezinger, S.~S. Ioushua, M.~Namer, Z.~Yang, and Y.~C. Eldar,
  ``{RF} chain reduction for {MIMO} systems: A hardware prototype,''
  \emph{{IEEE} Syst. J.}, 2020.

\bibitem{ioushua2019family}
S.~S. Ioushua and Y.~C. Eldar, ``A family of hybrid analog--digital beamforming
  methods for massive {MIMO} systems,'' \emph{{IEEE} Trans. Signal Process.},
  vol.~67, no.~12, pp. 3243--3257, 2019.

\bibitem{wang2019dynamic}
H.~Wang, N.~Shlezinger, Y.~C. Eldar, S.~Jin, M.~F. Imani, I.~Yoo, and D.~R.
  Smith, ``Dynamic metasurface antennas for {MIMO-OFDM} receivers with
  bit-limited {ADCs},'' \emph{{IEEE} Trans. Commun.}, 2020.

\bibitem{shlezinger2020dynamic}
N.~Shlezinger, G.~C. Alexandropoulos, M.~F. Imani, Y.~C. Eldar, and D.~R.
  Smith, ``Dynamic metasurface antennas for {6G} extreme massive {MIMO}
  communications,'' \emph{{IEEE} Wireless Commun. Mag.}, 2021.

\bibitem{liu2019low}
J.~Liu, Z.~Luo, and X.~Xiong, ``Low-resolution {ADCs} for wireless
  communication: A comprehensive survey,'' \emph{IEEE Access}, vol.~7, pp.
  91\,291--91\,324, 2019.

\bibitem{zhang2020deep}
Y.~Zhang, M.~Alrabeiah, and A.~Alkhateeb, ``Deep learning for massive {MIMO}
  with 1-bit {ADCs}: When more antennas need fewer pilots,'' 2020.

\bibitem{klautau2018detection}
A.~Klautau, N.~Gonz{\'a}lez-Prelcic, A.~Mezghani, and R.~W. Heath, ``Detection
  and channel equalization with deep learning for low resolution {MIMO}
  systems,'' in \emph{52nd Asilomar Conference on Signals, Systems, and
  Computers}.\hskip 1em plus 0.5em minus 0.4em\relax IEEE, 2018, pp.
  1836--1840.

\bibitem{balevi2019one}
E.~Balevi and J.~G. Andrews, ``One-bit {OFDM} receivers via deep learning,''
  \emph{{IEEE} Trans. Commun.}, vol.~67, no.~6, pp. 4326--4336, 2019.

\bibitem{balevi2019two}
------, ``Two-stage learning for uplink channel estimation in one-bit massive{
  MIMO},'' \emph{arXiv preprint arXiv:1911.12461}, 2019.

\bibitem{balevi2020autoencoder}
------, ``Autoencoder-based error correction coding for one-bit quantization,''
  \emph{{IEEE} Trans. Commun.}, 2020.

\bibitem{kim2019machine}
D.~Kim and N.~Lee, ``Machine learning based detections for mmwave two-hop
  {MIMO} systems using one-bit transceivers,'' in \emph{Proc. IEEE SPAWC},
  2019.

\bibitem{nguyen2020svm}
L.~V. Nguyen, A.~L. Swindlehurst, and D.~H. Nguyen, ``{SVM}-based channel
  estimation and data detection for one-bit massive {MIMO} systems,''
  \emph{arXiv preprint arXiv:2003.10678}, 2020.

\bibitem{nguyen2020linear}
------, ``Linear and deep neural network-based receivers for massive {MIMO}
  systems with one-bit {ADCs},'' \emph{arXiv preprint arXiv:2008.03757}, 2020.

\bibitem{choi2016near}
J.~Choi, J.~Mo, and R.~W. Heath, ``Near maximum-likelihood detector and channel
  estimator for uplink multiuser massive mimo systems with one-bit adcs,''
  \emph{{IEEE} Trans. Commun.}, vol.~64, no.~5, pp. 2005--2018, 2016.

\bibitem{risi2014massive}
C.~Risi, D.~Persson, and E.~G. Larsson, ``Massive {MIMO} with 1-bit {ADC},''
  \emph{arXiv preprint arXiv:1404.7736}, 2014.

\bibitem{jacobsson2015one}
S.~Jacobsson, G.~Durisi, M.~Coldrey, U.~Gustavsson, and C.~Studer, ``One-bit
  massive {MIMO}: Channel estimation and high-order modulations,'' in
  \emph{Proc. IEEE ICCW}, 2015, pp. 1304--1309.

\bibitem{ivrlac2007mimo}
M.~T. Ivrlac and J.~A. Nossek, ``On {MIMO} channel estimation with single-bit
  signal-quantization,'' in \emph{ITG smart antenna workshop}, 2007.

\bibitem{mo2014channel}
J.~Mo, P.~Schniter, N.~G. Prelcic, and R.~W. Heath, ``Channel estimation in
  millimeter wave mimo systems with one-bit quantization,'' in \emph{2014 48th
  Asilomar Conference on Signals, Systems and Computers}.\hskip 1em plus 0.5em
  minus 0.4em\relax IEEE, 2014, pp. 957--961.

\bibitem{mezghani2018blind}
A.~Mezghani and A.~L. Swindlehurst, ``Blind estimation of sparse broadband
  massive {MIMO} channels with ideal and one-bit {ADCs},'' \emph{{IEEE} Trans.
  Signal Process.}, vol.~66, no.~11, pp. 2972--2983, 2018.

\bibitem{li2017channel}
Y.~Li, C.~Tao, G.~Seco-Granados, A.~Mezghani, A.~L. Swindlehurst, and L.~Liu,
  ``Channel estimation and performance analysis of one-bit massive {MIMO}
  systems,'' \emph{IEEE Trans. Signal Process}, vol.~65, no.~15, pp.
  4075--4089, 2017.

\bibitem{jacobsson2017throughput}
S.~Jacobsson, G.~Durisi, M.~Coldrey, U.~Gustavsson, and C.~Studer, ``Throughput
  analysis of massive {MIMO} uplink with low-resolution {ADCs},'' \emph{{IEEE}
  Trans. Wireless Commun.}, vol.~16, no.~6, pp. 4038--4051, 2017.

\bibitem{mo2017channel}
J.~Mo, P.~Schniter, and R.~W. Heath, ``Channel estimation in broadband
  millimeter wave {MIMO} systems with few-bit {ADCs},'' \emph{{IEEE} Trans.
  Signal Process.}, vol.~66, no.~5, pp. 1141--1154, 2017.

\bibitem{lecun2015deep}
Y.~LeCun, Y.~Bengio, and G.~Hinton, ``Deep learning,'' \emph{Nature}, vol. 521,
  no. 7553, p. 436, 2015.

\bibitem{farsad2017detection}
N.~Farsad and A.~Goldsmith, ``Detection algorithms for communication systems
  using deep learning,'' \emph{arXiv preprint arXiv:1705.08044}, 2017.

\bibitem{corlay2018multilevel}
V.~Corlay, J.~J. Boutros, P.~Ciblat, and L.~Brunel, ``Multilevel {MIMO}
  detection with deep learning,'' in \emph{2018 52nd Asilomar Conference on
  Signals, Systems, and Computers}.\hskip 1em plus 0.5em minus 0.4em\relax
  IEEE, 2018, pp. 1805--1809.

\bibitem{liao2019deep}
Y.~Liao, N.~Farsad, N.~Shlezinger, Y.~C. Eldar, and A.~J. Goldsmith, ``Deep
  neural network symbol detection for millimeter wave communications,''
  \emph{arXiv preprint arXiv:1907.11294}, 2019.

\bibitem{shlezinger2019viterbinet}
N.~Shlezinger, N.~Farsad, Y.~C. Eldar, and A.~J. Goldsmith, ``{ViterbiNet}: A
  deep learning based {Viterbi} algorithm for symbol detection,'' \emph{{IEEE}
  Trans. Wireless Commun.}, vol.~19, no.~5, pp. 3319--3331, 2020.

\bibitem{shlezinger2020deepsic}
N.~Shlezinger, R.~Fu, and Y.~C. Eldar, ``{DeepSIC}: Deep soft interference
  cancellation for multiuser {MIMO} detection,'' \emph{{IEEE} Trans. Wireless
  Commun.}, 2020.

\bibitem{shlezinger2020data}
N.~Shlezinger, N.~Farsad, Y.~C. Eldar, and A.~J. Goldsmith, ``Data-driven
  factor graphs for deep symbol detection,'' in \emph{Proc. IEEE ISIT}, 2020.

\bibitem{he2019model}
H.~He, C.-K. Wen, S.~Jin, and G.~Y. Li, ``Model-driven deep learning for joint
  {MIMO} channel estimation and signal detection,'' \emph{arXiv preprint
  arXiv:1907.09439}, 2019.

\bibitem{samuel2019learning}
N.~Samuel, T.~Diskin, and A.~Wiesel, ``Learning to detect,'' \emph{{IEEE}
  Trans. Signal Process.}, vol.~67, no.~10, pp. 2554--2564, 2019.

\bibitem{takabe2019trainable}
S.~Takabe, M.~Imanishi, T.~Wadayama, R.~Hayakawa, and K.~Hayashi, ``Trainable
  projected gradient detector for massive overloaded mimo channels: Data-driven
  tuning approach,'' \emph{IEEE Access}, vol.~7, pp. 93\,326--93\,338, 2019.

\bibitem{balatsoukas2019deep}
A.~Balatsoukas-Stimming and C.~Studer, ``Deep unfolding for communications
  systems: A survey and some new directions,'' \emph{arXiv preprint
  arXiv:1906.05774}, 2019.

\bibitem{farsad2020data}
N.~Farsad, N.~Shlezinger, A.~J. Goldsmith, and Y.~C. Eldar, ``Data-driven
  symbol detection via model-based machine learning,'' \emph{arXiv preprint
  arXiv:2002.07806}, 2020.

\bibitem{hershey2014deep}
J.~R. Hershey, J.~L. Roux, and F.~Weninger, ``Deep unfolding: Model-based
  inspiration of novel deep architectures,'' \emph{arXiv preprint
  arXiv:1409.2574}, 2014.

\bibitem{monga2019algorithm}
V.~Monga, Y.~Li, and Y.~C. Eldar, ``Algorithm unrolling: Interpretable,
  efficient deep learning for signal and image processing,'' \emph{{IEEE}
  Signal Process. Mag.}, 2020.

\bibitem{khobahi2020unfolded}
N.~Naimipour, S.~Khobahi, and M.~Soltanalian, ``Unfolded algorithms for deep
  phase retrieval,'' \emph{arXiv preprint arXiv:2012.11102}, 2020.

\bibitem{agarwal2020deep}
C.~Agarwal, S.~Khobahi, A.~Bose, M.~Soltanalian, and D.~Schonfeld, ``Deep-url:
  A model-aware approach to blind deconvolution based on deep unfolded
  richardson-lucy network,'' in \emph{2020 IEEE International Conference on
  Image Processing (ICIP)}.\hskip 1em plus 0.5em minus 0.4em\relax IEEE, 2020,
  pp. 3299--3303.

\bibitem{khobahi2020deep}
S.~Khobahi, A.~Bose, and M.~Soltanalian, ``Deep radar waveform design for
  efficient automotive radar sensing,'' in \emph{2020 IEEE 11th Sensor Array
  and Multichannel Signal Processing Workshop (SAM)}.\hskip 1em plus 0.5em
  minus 0.4em\relax IEEE, 2020, pp. 1--5.

\bibitem{naimipour2020upr}
N.~Naimipour, S.~Khobahi, and M.~Soltanalian, ``Upr: A model-driven
  architecture for deep phase retrieval,'' \emph{arXiv preprint
  arXiv:2003.04396}, 2020.

\bibitem{shlezinger2020model}
N.~Shlezinger, J.~Whang, Y.~C. Eldar, and A.~G. Dimakis, ``Model-based deep
  learning,'' \emph{arXiv preprint arXiv:2012.08405}, 2020.

\bibitem{gregor2010learning}
K.~Gregor and Y.~LeCun, ``Learning fast approximations of sparse coding,'' in
  \emph{Proceedings of the 27th International Conference on Machine
  Learning}.\hskip 1em plus 0.5em minus 0.4em\relax Omnipress, 2010, pp.
  399--406.

\bibitem{flordelis2019massive}
J.~Flordelis, X.~Li, O.~Edfors, and F.~Tufvesson, ``Massive {MIMO} extensions
  to the {COST} 2100 channel model: Modeling and validation,'' \emph{{IEEE}
  Trans. Wireless Commun.}, vol.~19, no.~1, pp. 380--394, 2019.

\bibitem{kingma2014adam}
D.~P. Kingma and J.~Ba, ``Adam: A method for stochastic optimization,''
  \emph{arXiv preprint arXiv:1412.6980}, 2014.

\end{thebibliography}

\end{document}